%% file: 0_main.tex
\documentclass[%
 twocolumn,
 amsmath,amssymb,
 aps, 
 physrev,
 10pt,
 floatfix,
superscriptaddress,
nobibnotes,
]{revtex4-2}

\usepackage{graphicx}
\usepackage{xcolor}
\usepackage{dcolumn}
\usepackage{bm}
\usepackage{comment}
\usepackage{hyperref}
\usepackage[normalem]{ulem}
\hypersetup{
	colorlinks,
	linkcolor=[rgb]{0, 0, 0.8},
	citecolor=[rgb]{0, 0, 0.8},
	urlcolor=[rgb]{0, 0, 0.5}
}
\usepackage{booktabs}
\begin{document}

\newcommand{\deltar}{\delta_\mathrm{r}}
\newcommand{\gammarelax}{\gamma_-}
\newcommand{\gammadephasing}{\gamma_\phi}
\newcommand{\gammalead}{\Gamma_\mathrm{0e}}
\newcommand{\gammatot}{\Gamma_\mathrm{tot}}
\newcommand{\kdqd}{\kappa_\mathrm{DQD}}
\newcommand{\dir}{D}
\newcommand{\isd}{I_\mathrm{SD}}

\newcommand{\wj}[1]{\textcolor{blue}{WJ: #1}}
\newcommand{\fo}[1]{\textcolor{cyan}{FO: #1}}
\newcommand{\vm}[1]{\textcolor{red}{VM: #1}}

\include{1_paper}

\bibliography{0.1_references}


\end{document}

%% file: 1_paper.tex

\input{1.0_title_abstract}

\input{1.1_introduction}
\input{1.2_device}

\input{1.3_photodetection}

\input{1.4_efficiency}

\input{1.5_tunability}

\input{1.6.1_discussion}

\appendix
\input{2_appendix}


%% file: 1.0_title_abstract.tex

\title{\textbf{High-Efficiency Tunable Microwave Photon Detector Based on a Semiconductor Double Quantum Dot Coupled to a Superconducting High-Impedance Cavity} 
}%

\newcommand{\epfl}{Hybrid Quantum Circuits Laboratory, Institute of Physics, École Polytéchnique Fédérale de Lausanne (EPFL), 1015 Lausanne, Switzerland}
\newcommand{\epflqse}{Hybrid Quantum Circuits Laboratory, Center for Quantum Science and Engineering, École Polytéchnique Fédérale de Lausanne (EPFL), 1015 Lausanne, Switzerland}
\newcommand{\basel}{Department of Physics, University of Basel, Klingelbergstrasse 82, 4056 Basel, Switzerland}
\newcommand{\ethz}{Laboratory for Solid State Physics, ETH Zürich, 8093 Zürich, Switzerland}
\newcommand{\ethzqc}{Quantum Center, ETH Zürich, 8093 Zürich, Switzerland}
\newcommand{\lund}{NanoLund and Solid State Physics, Lund University, Box 118, 22100 Lund, Sweden}

\author{Fabian Oppliger}%
\thanks{These authors contributed equally to this work.}
\author{Wonjin Jang}%
\thanks{These authors contributed equally to this work.}
\affiliation{\epfl}%
\affiliation{\epflqse}%

\author{Aldo Tarascio}%
\affiliation{\basel}%

\author{Franco De Palma}%
\affiliation{\epfl}%
\affiliation{\epflqse}%

\author{Christian Reichl}%
\affiliation{\ethz}%
\affiliation{\ethzqc}%

\author{Werner Wegscheider}%
\affiliation{\ethz}%
\affiliation{\ethzqc}%

\author{Ville F. Maisi}%
\affiliation{\lund}%

\author{Dominik Zumbühl}%
\affiliation{\basel}%

\author{Pasquale Scarlino}
\email[Contact author: ]{pasquale.scarlino@epfl.ch}
\affiliation{\epfl}%
\affiliation{\epflqse}%

\date{\today}

\begin{abstract}
High-efficiency single-photon detection in the microwave domain is a key enabling technology for quantum sensing, communication, and information processing.
However, the extremely low energy of microwave photons ($\sim\mu$eV) presents a fundamental challenge, preventing direct photon-to-charge conversion as achieved in optical systems using semiconductors.
Semiconductor quantum dot (QD) charge qubits offer a compelling solution due to their highly tunable energy levels in the microwave regime, enabling coherent coupling with single photons.
In this work, we demonstrate microwave photon detection with an efficiency approaching 70\% in the single-photon regime.
We use a hybrid system comprising a double quantum dot (DQD) charge qubit electrostatically defined in a GaAs/AlGaAs heterostructure,
coupled to a high-impedance Josephson junction (JJ) array cavity.
We systematically optimize the hybrid device architecture to maximize the  conversion efficiency,
leveraging the strong charge--photon coupling and the tunable DQD tunnel coupling rates.
Incoming cavity photons coherently excite the DQD qubit,
which in turn generates a measurable electrical current, realizing deterministic photon-to-charge conversion.
Moreover, by exploiting the independent tunability of both the DQD transition energy and the cavity resonance frequency,
we characterize the system efficiency over a range of $3-5.2$~GHz.
Our results establish semiconductor-based cavity-QED architectures as a scalable
and versatile platform for efficient microwave photon detection,
opening new avenues for quantum microwave optics and hybrid quantum information technologies.

\end{abstract}

\maketitle


%% file: 1.1_introduction.tex

\section{Introduction}\label{sec:intro}

The ability to efficiently detect single photons is fundamental to a broad range of quantum technologies, including quantum communication, cryptography, sensing, and information processing \cite{hadfield_single-photon_2009, gisin_quantum_2002, obrien_optical_2007, hadfield_single-photon_2023}. 
While in the optical domain single-photon detection is a mature technology—with photodiodes, avalanche photodetectors, and superconducting nanowires achieving near-unity efficiency \cite{hadfield_single-photon_2009, eisaman_invited_2011, cova_towards_1981, dautet_photon_1993, takesue_quantum_2007}—the transition to the microwave regime presents profound challenges. 
At gigahertz frequencies, the energy of a single photon is five orders of magnitude smaller than that of an optical photon, rendering traditional photoelectric detection mechanisms inapplicable \cite{gu_microwave_2017, sathyamoorthy_detecting_2016}. 
As a result, alternative architectures are required to realize microwave single-photon detectors capable of operating at the quantum limit.

Various strategies have been developed to address this challenge,
leveraging tools from the circuit quantum electrodynamics (cQED) framework \cite{gu_microwave_2017}. 
These include superconducting qubit-based detectors that map photon absorption onto quantum state transitions
\cite{balembois_cyclically_2024, besse_single-shot_2018, inomata_single_2016},
biased Josephson junctions
\cite{chen_microwave_2011, stanisavljevic_efficient_2024, pankratov_detection_2025, chai_measuring_2025},
parametrically driven Kerr resonators operated at criticality \cite{petrovnin_microwave_2024}
and bolometric sensors based on materials such as graphene
\cite{lee_graphene-based_2020, kokkoniemi_bolometer_2020, chang_quantum-ready_2025}. 
While these approaches have demonstrated key milestones such as quantum non-demolition detection and sub-femtowatt sensitivity,
they are often constrained by requirements for pulsed operation, complex reset protocols, sensitivity to quasiparticles,
or limited long-term stability
\cite{balembois_cyclically_2024, besse_single-shot_2018, inomata_single_2016, chen_microwave_2011, stanisavljevic_efficient_2024, pankratov_detection_2025, chai_measuring_2025}. 
Moreover, continuous and passive detection of itinerant microwave photons remains elusive.

An emerging and promising approach involves hybrid semiconductor-superconductor architectures,
in particular double quantum dots (DQDs) embedded in high-impedance microwave cavities
\cite{stockklauser_strong_2017, mi_strong_2017, de_palma_strong_2024, samkharadze_strong_2018}. 
When operated in the charge qubit regime, DQDs possess energy splittings in the microwave domain,
allowing them to absorb cavity photons and convert them into a measurable electrical current—a mechanism analogous to that of optical photodiodes 
\cite{khan_efficient_2021, wong_quantum_2017, haldar_continuous_2024, zenelaj_wigner-function_2025}. 
Unlike many superconducting qubit-based detectors, DQD detectors can operate continuously,
without active qubit reset or prior knowledge of photon arrival time
\cite{khan_efficient_2021, haldar_continuous_2024, haldar_high-efficiency_2024, zenelaj_wigner-function_2025}. 
Their well-defined energy structure provides intrinsic frequency selectivity,
and the absence of Josephson junctions in the absorption medium eliminates sensitivity to quasiparticle-induced dark counts,
a key advantage for robust operation \cite{chen_microwave_2011, stanisavljevic_efficient_2024, pankratov_detection_2025, chai_measuring_2025}.

In recent years, pioneering experiments have demonstrated photon-assisted tunneling (PAT) in DQD–cavity hybrid systems,
validating their potential as microwave photodetectors \cite{khan_efficient_2021, haldar_continuous_2024, haldar_high-efficiency_2024}. 
However, the reported efficiencies have so far remained limited, largely due to the poor tunability of the semiconducting host materials
used and the low impedance of the superconducting cavities employed.
These limitations have constrained the coupling strength and matching of photon absorption rates, necessary for optimal photoconversion.
As a result, the realization of high-efficiency, continuous microwave photon detection in such systems has remained an open challenge.

In this work, we demonstrate a semiconductor DQD-based cavity detector operating in a fully passive and continuous mode.
We report a microwave photon detection efficiency approaching 70\%, marking a significant improvement
over previously reported semiconductor-based detectors \cite{khan_efficient_2021, haldar_high-efficiency_2024}.
This performance is enabled by strong charge-photon coupling, made possible through the high impedance of the superconducting cavity,
the \textit{in-situ} tunability of the gate-defined DQD tunneling rates, as well as the inherently low charge relaxation rate of the DQD.
In addition, the frequency tunability of both the cavity and the DQD excitation energy
allows for frequency-selective detection across a broad range of $3-5.2$~GHz.
To accurately quantify the detection efficiency in the single-photon regime,
we perform a careful calibration of the input photon flux via ac Stark shift measurements of the charge qubit frequency~\cite{schuster_ac_2005}.
Together, these results mark a crucial step toward scalable and high-fidelity microwave photon detection in hybrid quantum systems.

%% file: 1.2_device.tex

\section{Device Architecture}\label{sec:device}

\begin{figure}[!ht]
    \includegraphics[width=\linewidth]{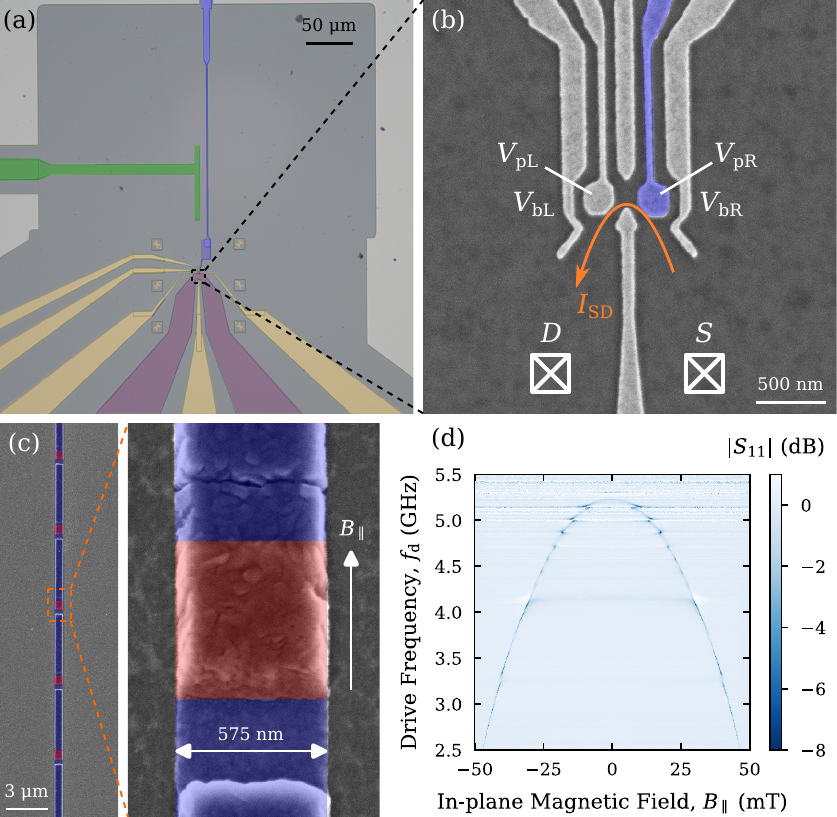}
	\caption{Hybrid device architecture and frequency tunability of the microwave cavity.
		(a) False-colored optical micrograph of the hybrid device with the double quantum dot (DQD) gates in yellow, the mesa region with the 2DEG in purple,
        the Josephson junction (JJ) array cavity in blue and the coupling capacitor to the feedline in green.
		(b) Scanning electron micrograph of the DQD gate structure. The right plunger gate (in blue) is galvanically connected to the end of the JJ array cavity.
		(c) Scanning electron micrograph of the JJ array cavity with a zoom-in on a single junction. The regions containing the JJs are highlighted in red.
        (d) Normalized cavity reflectance $|S_{11}|$ measured at low drive power as a function of the cavity drive frequency $f_\mathrm{d}$
        and in-plane magnetic field $B_\parallel$ parallel to the JJ array, indicated by the white arrow in (c).
		The cavity resonance frequency is tuned as the Josephson inductance $L$ is modulated by the magnetic flux penetrating the JJs \cite{kuzmin_tuning_2023}.
	}
    \label{fig:fig1}
\end{figure}

Figure~\ref{fig:fig1}(a) shows an optical micrograph of the semiconductor-superconductor hybrid circuit quantum electrodynamics (cQED) device built on a GaAs/AlGaAs heterostructure.
A 2-dimensional electron gas (2DEG) is accumulated at the top GaAs/AlGaAs interface via remote doping.
To form a well-defined conductive channel and to minimize dissipation of microwave signals,
the 2DEG is locally removed by etching, such that only a small mesa region remains
(purple shaded region in Fig.~\ref{fig:fig1}(a)).
Ohmic contacts to the 2DEG (white crossed boxes in Fig.~\ref{fig:fig1}(b)) allow to probe the source-drain current
$\isd$ across the DQD (see setup described in Appendix~\ref{sec:app_setup}).
A scanning electron micrograph of a zoomed-in region denoted by the dashed box in Fig.~\ref{fig:fig1}(a) is shown in Fig.~\ref{fig:fig1}(b).
A DQD is laterally defined by Ti/Au gates patterned on top of the heterostructure (Fig.~\ref{fig:fig1}(b)).
The right plunger gate, shaded in blue in Fig.~\ref{fig:fig1}(b), is galvanically connected to a superconducting high-impedance cavity.
The cavity consists of an array of $N$=24 Al/Al$_2$O$_3$/Al Josephson junctions (JJs) and is shown in Fig.~\ref{fig:fig1}(c) (blue shaded structure in Fig.~\ref{fig:fig1}(a)) \cite{scarlino_situ_2022}.
The other end of the cavity is shunted to ground to form a quarter-wave cavity.
The JJ array is fabricated by two angled evaporation steps with an oxidation step in between to form a JJ in the overlapping region of the two Al electrodes (red shaded region in Fig.~\ref{fig:fig1}(c)).
Further fabrication details can be found in Appendix~\ref{sec:app_fab}.

A single-port feedline is capacitively coupled to the cavity (green shaded structure in Fig.~\ref{fig:fig1}(a)), allowing to probe the cavity response with a reflection measurement.
The microwave photons transferred from the feedline to the cavity are absorbed by the DQD enabling photon-to-electron conversion \cite{khan_efficient_2021, haldar_high-efficiency_2024}. 
Thereby, it is preferable to maximize the feedline-cavity coupling rate $\kappa_\mathrm{c}$ to enable efficient photon transfer,
as will be detailed later.
Notably, the resonance frequency $f_\mathrm{c}$ of the JJ array cavity is tunable by applying a magnetic field $B_\parallel$ parallel to the JJ array
\cite{kuzmin_tuning_2023} as demonstrated in Fig.~\ref{fig:fig1}(d). 
At $B_\parallel=0$, the cavity resonates at $f_\mathrm{c}\sim 5.2$~GHz,
set by the total equivalent inductance ($L \sim 36$ nH), and the total stray capacitance ($C \sim 26$ fF) of the cavity,
resulting in an impedance of $Z \sim 1.2~\mathrm{k}\Omega$ \cite{scarlino_situ_2022}.
A finite $B_\parallel$ penetrating the JJs effectively tunes $L$ to induce a modulation of the resonance frequency \cite{kuzmin_tuning_2023}.
We attribute the spurious resonances that appear independently of $B_\parallel$ in Fig.~\ref{fig:fig1}(d)
to parasitic modes arising from the device packaging~\cite{huang_microwave_2021},
insufficient chip grounding (leading to slot modes)~\cite{chen_fabrication_2014, ponchak_excitation_2005},
and impedance mismatch introduced by the limited bandwidth of the circulators above 4.5\,GHz.
The latter is mitigated by using a different circulator setup for high-frequency measurements
(see Appendix~\ref{sec:app_setup}).

%% file: 1.3_photodetection.tex
%
\section{Photodetection}\label{sec:photodetection}
\begin{figure}[!ht]
    \includegraphics[width=0.95\linewidth]{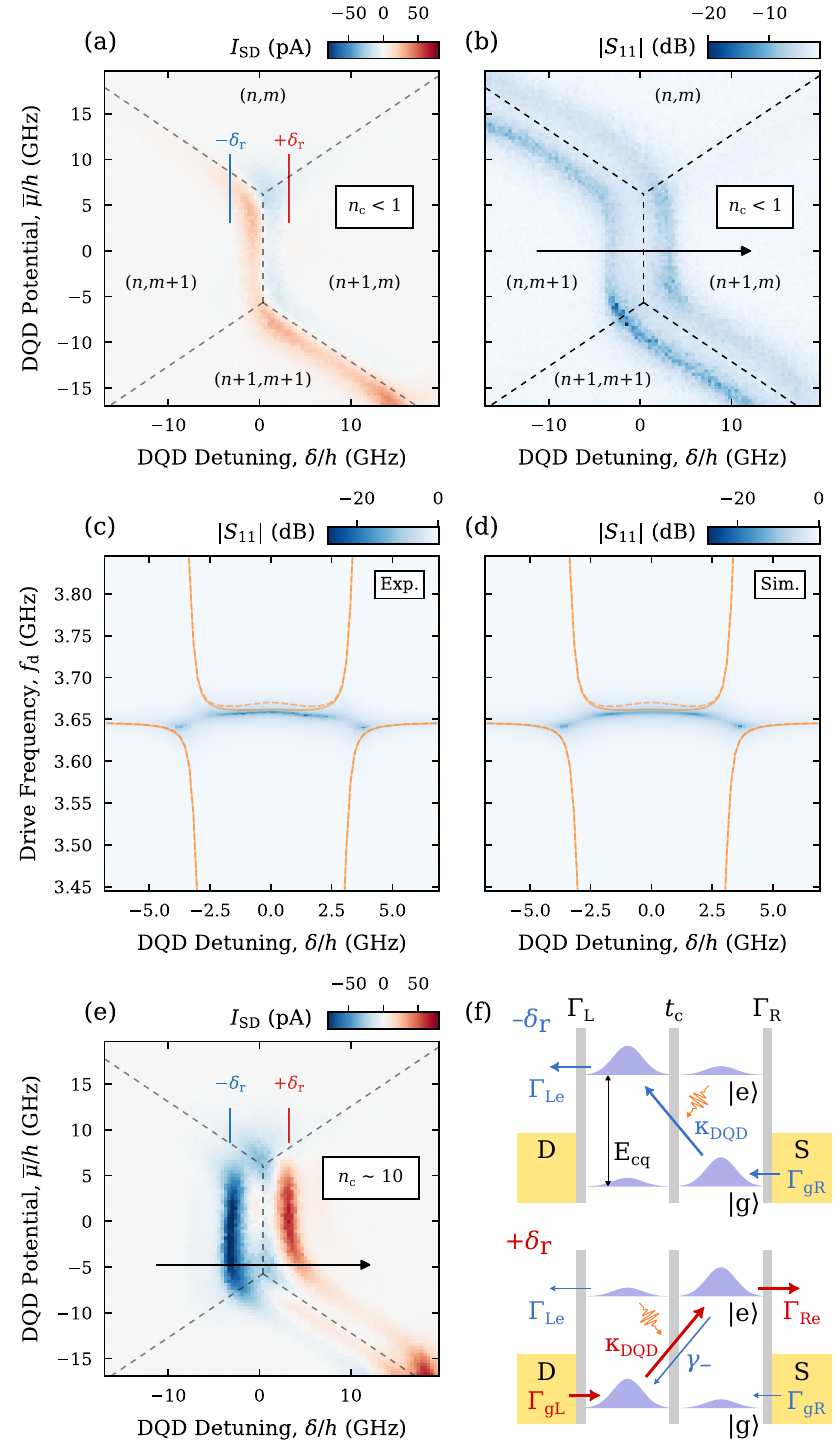}
	\caption{Observation of photon-induced dc-current.
        (a-b) DQD charge stability diagram as a function of DQD detuning $\delta$ and potential $\bar{\mu}$
            recorded by  simultaneously monitoring (a) the DQD source-drain current $\isd$ and (b) the normalized cavity reflectance $|S_{11}|$,
            measured with a low drive power ($n_\mathrm{c} < 1$).
            $\pm \deltar$ corresponds to the DQD detuning at which the charge qubit energy matches that of the cavity photon.
            $(n,m)$ represents the charge occupation numbers of the DQD.
        (c) $|S_{11}|$ measured as a function of drive frequency $f_\mathrm{d}$ and $\delta$,
            with the bare cavity resonance frequency $f_\mathrm{c} = 3.646$~GHz
            (d) Simulation of $|S_{11}|$ with the parameters extracted from a numerical fit of (c) to an input-output model
            built from a full Rabi model Hamiltonian (see Appendix~\ref{sec:app_io}). 
            For (c-d), solid (dashed) orange curve corresponds to eigenspectrum derived from a full Rabi (Jaynes-Cummings) Hamiltonian.
		(e) DQD stability diagram in the same region as panel (a), but measured with large cavity drive power ($n_\mathrm{c} \sim 10$).
		Additional non-zero current occurs at $\delta=\pm\deltar$, indicating the presence of photon-assisted tunneling (PAT).
		(f) Schematic explaining the photoconversion process for $\delta=+\deltar$ (top panel) and $\delta=-\deltar$ (bottom panel).
        For clarity, the unwanted quenching processes $\gamma_-$, $\Gamma_\mathrm{Re}$ and $\Gamma_\mathrm{gL}$ are omitted in the top panel.
	}
    \label{fig:fig2}
\end{figure}

We first present the charge stability diagram of the DQD in Fig.~\ref{fig:fig2}(a),
measured by recording the source-drain current $\isd$ as a function of the average chemical potential $\bar{\mu} = (\mu_\mathrm{L} + \mu_\mathrm{R})/2$
and the inter-dot detuning $\delta = \mu_\mathrm{R} - \mu_\mathrm{L}$, with zero bias applied between the source and drain contacts (see Fig.~\ref{fig:fig2}(f)).
Here, $\mu_\mathrm{L}$ and $\mu_\mathrm{R}$ denote the chemical potentials of the left and right quantum dots, respectively.
$\bar{\mu}$ and $\delta$ are tuned individually by virtualizing the barrier gate voltages $V_\mathrm{bL}$ and $V_\mathrm{bR}$ (see Fig.~\ref{fig:fig1}(b)) \cite{taubert_determination_2011}.
In this configuration, the DQD is tuned to have fast tunneling rates to the left ($\Gamma_\mathrm{L}$) and right ($\Gamma_\mathrm{R}$) reservoirs
while maintaining a low inter-dot tunneling rate $t_\mathrm{c}$ relative to the cavity frequency.
$(n,m)$ in Fig.~\ref{fig:fig2}(a) represents the charge occupation numbers of the left and right dot, respectively.
In this configuration, we are able to simultaneously record $\isd$ and the cavity response.

Due to the large tunnel coupling to the reservoirs, we observe clear signatures of cotunneling current between the right dot and the right reservoir.
We attribute the antisymmetric currents around $\delta = 0$ to gate-voltage-induced noise,
predominantly from the left plunger gate $V_\mathrm{pL}$ (Fig.~\ref{fig:fig1}(b)),
likely resulting from suboptimal attenuation in the microwave line~\cite{entin-wohlman_heat_2017} (see Appendix~\ref{sec:app_noise} for discussion).
However, these currents vanish at finite detunings $\delta = \pm\deltar$ (indicated by the red and blue lines in Fig.~\ref{fig:fig2}(a)),
where the detector operates, and therefore do not affect the photon detection measurements discussed later in this work.
$\delta = \pm\deltar = \pm\sqrt{(hf_\mathrm{c})^2 - 4t_\mathrm{c}^2}$,
corresponds to the DQD detuning at which the DQD charge qubit energy $E_\mathrm{cq} = \sqrt{\delta^2 + 4t_\mathrm{c}^2}$ is resonant with the cavity mode.

The normalized cavity reflectance $|S_{11}|$ measured at resonance is presented in Fig.~\ref{fig:fig2}(b),
with the cavity resonance frequency tuned to $f_\mathrm{c} = 3.646$~GHz.
Initially, to avoid exciting the charge qubit, we use a low drive power such that the average intra-cavity photon number $n_\mathrm{c}$ remains well below unity.
In Fig.~\ref{fig:fig2}(b), dips in $|S_{11}|$ appear whenever the cavity interacts with dissipative transitions in the DQD system.
These features indicate an increase in the internal cavity loss rate $\kappa_\mathrm{i}$,
which enhances the depth of the resonance dip of an overcoupled cavity measured in reflection~\cite{scigliuzzo_effects_2021}.
Transitions involving the right QD and its reservoir, $(n,m) \leftrightarrow (n,m+1)$ and $(n+1,m) \leftrightarrow (n+1,m+1)$, are clearly visible.
This is consistent with the cavity being directly coupled, and therefore more sensitive,
to the right plunger gate (see Fig.~\ref{fig:fig1}(b)).
The observed broadening of these transitions is attributed to the large tunnel coupling between the right QD and its reservoir.

We further elucidate the hybridization between the DQD and the cavity by monitoring the cavity reflectance as a function of the charge qubit energy,
varying $\delta$ along the black arrow in Fig.~\ref{fig:fig2}(b)
while keeping the inter-dot tunnel coupling $t_\mathrm{c}$ fixed.
Figure~\ref{fig:fig2}(c) displays the measured cavity spectroscopy,
obtained by recording the normalized reflectance $|S_{11}|$ as a function of both $\delta$ and the cavity drive frequency $f_\mathrm{d}$,
with the average DQD chemical potential $\bar{\mu} = 0$ and average intra-cavity photon number $n_\mathrm{c} < 1$.
Taking advantage of the frequency tunability of our cavity, we repeat the spectroscopy at different $f_\mathrm{c}$
and fit all the collected spectra simultaneously using an input–output model that includes counter-rotating terms of the full Rabi Hamiltonian~\cite{forn-diaz_observation_2010} 
(see Appendix~\ref{sec:app_io} for additional measurements and fitting details).
It is important to include these counter-rotating terms, since $g/(2\pi f_\mathrm{q})$
reaches values on the order of 0.1 at $\delta = 0$~\cite{scarlino_situ_2022, forn-diaz_observation_2010}.
From the fit, we extract a charge–photon coupling strength of $g_0/2\pi \sim 213.7 \pm 0.3$~MHz
(see Table~\ref{tab:app_parameters} for all extracted parameters).
Evidently, the eigenspectrum calculated with a simple Jaynes-Cummings model (dashed curves in Figs.~\ref{fig:fig2}(c) and (d))
does not closely capture the Bloch-Siegert shift near $\delta = 0$,
which is present in the experiment \cite{forn-diaz_observation_2010, niemczyk_circuit_2010, devoret_circuit-qed_2007}.
Instead, this shift is better reproduced by the full Rabi model (solid curves in Figs.~\ref{fig:fig2}(c) and (d)),
implying that the large charge-photon coupling strength significantly alters the dynamics of the hybrid system
\cite{forn-diaz_observation_2010, niemczyk_circuit_2010, devoret_circuit-qed_2007}.

As discussed in more detail later, cotunneling processes play a notable role in this configuration
due to the large tunnel coupling between the QDs and the reservoirs~\cite{de_franceschi_electron_2001, amasha_pseudospin-resolved_2013}.
Even when the excited state of the charge qubit remains below the reservoir Fermi level ($\bar{\mu} = 0$ and $\delta \ll U$,
where $U$ is the inter-dot charging energy), electrons can tunnel out via cotunneling~\cite{de_franceschi_electron_2001, amasha_pseudospin-resolved_2013}.
As a result, the total decoherence rate extracted from the input–output model,
$\gammatot/2\pi = (\gammalead + \gammarelax + 2\gammadephasing)/2\pi = 829.3 \pm 3.6$~MHz,
encompasses not only the intrinsic inter-dot relaxation and dephasing rates ($\gammarelax$ and $\gammadephasing$, respectively)
but also the relaxation to the reservoirs $\gammalead$ induced by cotunneling.

After having characterized our charge-photon hybrid system at low drive ($n_\mathrm{c}<1$),
in Fig.~\ref{fig:fig2}(e) we increase the drive power to populate the cavity with $n_\mathrm{c} \sim 10$
and measure the same region of the DQD stability diagram reported in Fig.~\ref{fig:fig2}(a).
Two distinct lines with positive and negative current appear exactly at $\delta=\pm\deltar$ (denoted by the red and blue lines),
where the qubit and the cavity are in resonance ($f_\mathrm{q}=E_\mathrm{cq}/h=f_\mathrm{c}$),
indicating the presence of photo-induced currents.
As depicted in the top panel of Fig.~\ref{fig:fig2}(f), at $\delta=-\deltar$, the qubit ground (excited) state is mostly localized in the right (left) dot.
An incoming photon, whose energy matches that of the qubit, can be absorbed with rate $\kdqd=4g^2/\gammatot$ and excite the qubit.
Due to the spatial structure of the ground and excited state wavefunctions, this essentially represents a photo-induced inter-dot tunneling event,
a process known as photon-assisted tunneling (PAT) \cite{oosterkamp_photon-assisted_1996, fujisawa_photon_1997}.
From the excited state, the electron can then tunnel out to the left reservoir with rate 
$\Gamma_\mathrm{Le}=\Gamma_\mathrm{L}\sin^2(\theta/2)$
and the ground state can be recovered by an electron tunneling in from the right reservoir with rate 
$\Gamma_\mathrm{gR}=2\Gamma_\mathrm{R}\sin^2(\theta/2)$,
where $\theta = \arccos(-\deltar/hf_\mathrm{c})$ is the DQD mixing angle.
The factor 2 in 
$\Gamma_\mathrm{gR}$ takes into account spin degeneracy \cite{haldar_high-efficiency_2024}.
Analogously, for $\delta=+\deltar$ (bottom panel of Fig.~\ref{fig:fig2}(f)),
the PAT process enables the transport of electrons from the left to the right reservoir with the rates $\Gamma_\mathrm{Re}=\Gamma_\mathrm{R}\cos^2(\theta/2)$
and $\Gamma_\mathrm{gL}=2\Gamma_\mathrm{L}\cos^2(\theta/2)$ instead.
When continuously driving the cavity, this can be observed as a positive (negative) current measured across the DQD at $+\deltar$ ($-\deltar$).
In the bottom panel of Fig.~\ref{fig:fig2}(f), further processes that disrupt the current across the DQD
($\gammarelax$, $\Gamma_\mathrm{Le}$ and $\Gamma_\mathrm{gR}$) are depicted,
which will be discussed in more detail in the next section.

While photo-induced currents typically appear only at $\pm\deltar$ near the charge triple points~\cite{khan_efficient_2021, haldar_high-efficiency_2024},
in the DQD configuration investigated here, such currents are observed as vertical lines throughout the entire region between the QD–reservoir transitions,
as shown in Fig.~\ref{fig:fig2}(e).
This stems from the aforementioned large cotunneling rate, enabled by the highly transparent tunnel barriers between the QDs and the reservoirs,
as well as a relatively small inter-dot charging energy $U$~\cite{amasha_pseudospin-resolved_2013}. 
In contrast, for an alternative DQD configuration (see Fig.~\ref{fig:suppfig_other_inter-dot} in Appendix~\ref{sec:app_other_inter-dot})
measured within the same device---featuring less transparent QD-reservoir barriers and a larger $U$---cotunneling processes
are significantly reduced~\cite{amasha_pseudospin-resolved_2013}.
As a result, PAT currents are confined primarily to the vicinity of the charge triple points.

%% file: 1.4_efficiency.tex

\section{Near-Unity Photon Detection Efficiency}\label{sec:efficiency}

\begin{figure*}[!ht]
    \includegraphics[width=\linewidth]{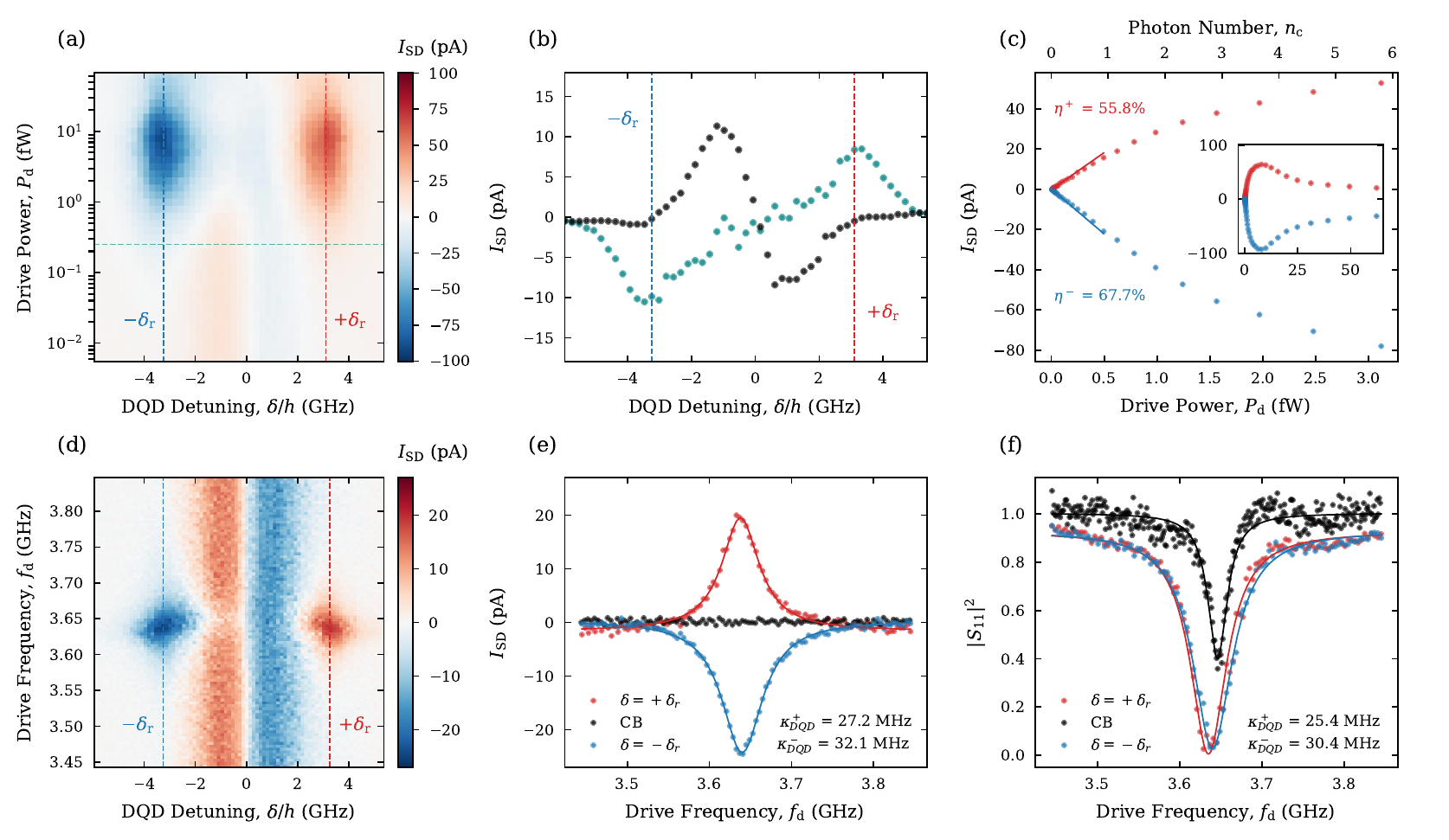}
	\caption{Photon detection efficiency and photon absorption rate.
		(a) DQD source-drain current $\isd$ measured as a function of feedline drive power $P_\mathrm{d}$ and DQD detuning $\delta$ near the charge triple points.
		The red (blue) dashed line indicates where $\delta=+\deltar$ ($\delta=-\deltar$).
        (b) The black dots denote the horizontal line cut of panel (a) at low $P_\mathrm{d} < 0.01$~fW.
        The cyan dots represent the effective photocurrent at finite $P_\mathrm{d}$
        obtained by taking a line cut of panel (a) at the cyan dashed line and subtracting the current at low $P_\mathrm{d}$ (black dots).
        The photocurrent features clear extrema at $\delta=\pm\deltar$ (red and blue dashed lines)
        demonstrating the photon-assisted tunneling process.
	(c) Vertical line cuts of panel (a) at $\delta=\pm\deltar$ for $P_\mathrm{d} < 3.5$~fW.
        A linear fit at low power ($0 < n_\mathrm{c} < 1$) yields a photon detection efficiency of
        $\eta=\isd/e\dot{N} = 55.8 \pm 4.0\%\ (67.7 \pm 4.8\%)$ for $+\deltar$ ($-\deltar$).
        The average intra-cavity photon number $n_\mathrm{c}$ is indicated in the top axis.
        The inset shows the same line cuts for the full measured range in $P_\mathrm{d}$.
        (d) Measured $\isd$ as a function of cavity drive frequency $f_\mathrm{d}$ and $\delta$ at $n_\mathrm{c}\sim 1$.
        (e) Line cut of panel (d) at $\delta=\pm\deltar$ and in Coulomb blockade (CB).
        From a Lorentzian fit to the measured $\isd$, represented by solid lines,
        we extract the total linewidth $\kappa_\mathrm{tot}=\kappa+\kdqd$.
        Together with the bare cavity linewidth $\kappa=28.1 \pm 0.3$~MHz,
        we obtain the DQD photon absorption rates $\kdqd^+=27.2 \pm 1.3$~MHz and $\kdqd^-=32.1 \pm 1.5$~MHz
        for $+\deltar$ and $-\deltar$, respectively.
        (f) Normalized cavity reflection $|S_{11}|^2$ as a function of $f_\mathrm{d}$,
        with the DQD in CB and at $\delta=\pm\deltar$, measured simultaneously with panel (e).
        The solid lines represent a fit to the input-output model in Eq.~\ref{eq:io}, from which we extract 
        $\kdqd^+=25.4 \pm 3.0$~MHz and $\kdqd^-=30.4 \pm 2.9$~MHz.
	}
    \label{fig:fig3}
\end{figure*}

To quantify the performance of this device as a photodetector, we measure the photon detection efficiency
$\eta = \isd/e\dot{N}$, where $\dot{N} = P_\mathrm{d}/hf_\mathrm{c}$ is the photon flux into the cavity feedline
and $P_\mathrm{d}$ is the feedline drive power at the coupling capacitor of the cavity port.
$P_\mathrm{d}$ is calibrated by measuring the ac Stark shift \cite{schuster_ac_2005} of the charge qubit frequency
(see Appendix~\ref{sec:app_ac-stark} for the full calibration procedure).
In Fig.~\ref{fig:fig3}(a), we show the measured $\isd$ as a function of $P_\mathrm{d}$ while varying the DQD detuning $\delta$
across the inter-dot transition in Fig.~\ref{fig:fig2}(e) close to the charge triple point (indicated by the black arrow).
The black dots in Fig.~\ref{fig:fig3}(b) shows a horizontal line cut from Fig.~\ref{fig:fig3}(a) obtained at low $P_\mathrm{d}<0.01$~fW
representing the dark current of this specific configuration.
The cyan dots in Fig.~\ref{fig:fig3}(b) represent a line cut at finite power after subtracting this dark current,
indicated by the cyan dashed line in Fig.~\ref{fig:fig3}(a).
The photocurrent features clear extrema at $\delta=\pm\deltar$ (red and blue dashed lines in Fig.~\ref{fig:fig3}(b)) demonstrating the PAT process.
Figure.~\ref{fig:fig3}(c) shows vertical line cuts of Fig.~\ref{fig:fig3}(a) at $\delta=\pm\deltar$ after subtracting the dark current at $\pm\deltar$.
For low $P_\mathrm{d}$ ($n_\mathrm{c} < 1$), the PAT current $\isd$ increases linearly,
as expected from the master equation model of the system \cite{khan_efficient_2021, haldar_continuous_2024}.
For $n_\mathrm{c} > 1$, nonlinear effects due to multi-photon processes
\cite{khan_efficient_2021, haldar_continuous_2024, haldar_high-efficiency_2024}
as well as the self-Kerr nonlinearity of the JJ array cavity (see Appendix~\ref{sec:app_kappas}) \cite{eichler_controlling_2014}
begin to play a significant role, hindering the photoconversion process.
A linear fit to the slope in the range of $0 < n_\mathrm{c} < 1$ yields an efficiency of $\eta^+=55.0 \pm 4.0\%$ and $\eta^-= 67.7 \pm 4.8\%$
at $+\deltar$ and $-\deltar$, respectively.

We now fix $P_\mathrm{d}$ and investigate the spectral response of the photodetector,
by measuring $\isd$ as a function of $\delta$ and the drive frequency $f_\mathrm{d}$ at $n_\mathrm{c}\sim 1$, as shown in Fig.~\ref{fig:fig3}(d).
As expected from theory \cite{khan_efficient_2021}, the photocurrent is maximized when driving at resonance, i.e. $f_\mathrm{d} = f_\mathrm{q} = f_\mathrm{c}$,
which only occurs at $\delta=\pm\deltar$.
Around these points, the model predicts a Lorentzian line-shape with a linewidth of $\kappa_\mathrm{tot} = \kappa + \kdqd$,
where $\kappa = \kappa_\mathrm{c} + \kappa_\mathrm{i}$ is the total bare resonator linewidth
and $\kdqd=4g^2/\gammatot$ can be interpreted as the effective photon absorption rate of the DQD \cite{khan_efficient_2021, wong_quantum_2017, haldar_high-efficiency_2024}.
Fig.~\ref{fig:fig3}(e) shows a Lorentzian fit to a cut from Fig.~\ref{fig:fig3}(d) at $+\deltar$ ($-\deltar$),
which yields $\kdqd^+=27.3\pm 1.5$~MHz ($\kdqd^-=32.1\pm 1.3$~MHz).
Another cut at large detuning $\delta>\deltar$ (black dots in Fig.~\ref{fig:fig3}(e)), for which the DQD is well in Coulomb blockade (CB),
confirms that the photon absorption process is quenched when the two systems are far detuned.
Alternatively, we can investigate the photon absorption by looking directly at the cavity signal instead.
In Fig.~\ref{fig:fig3}(f), we report the normalized cavity reflectance $|S_{11}|^2$ at $\delta=\pm\deltar$
and in CB, measured simultaneously with the $\isd$ shown in Fig.~\ref{fig:fig3}(e).
Here, the photon absorption by the DQD can be observed as an additional broadening of the resonance dip.
Fitting the input-output model in Eq.~(\ref{eq:io}) to the measured data
and subtracting the bare cavity linewidth from the $\kappa_\mathrm{tot}$ yields $\kdqd^+=25.4\pm 3.0$~MHz and $\kdqd^-=30.4\pm 2.9$~MHz,
which are close to the values previously extracted.
This further confirms that the measured current through the DQD is indeed induced by absorbing photons from the cavity.

In order to understand what limits the photon detection efficiency from reaching unity in the current device,
we use a master equation model to describe the induced photocurrent of such a hybrid system \cite{wong_quantum_2017, khan_efficient_2021}.
In the low-drive limit, the induced photocurrent can be described with a linear dependence
on the incoming photon flux $\dot{N}=P_\mathrm{d}/hf_\mathrm{c}$ as follows:
\begin{equation}
    \isd/e = \dot{N}
    \frac{\kappa_c}{\kappa}
    \frac{4\kdqd\kappa}{(\kdqd + \kappa)^2}
    \frac{\gammalead}{\gammalead + \gammarelax}
    \dir.
    \label{eq:photocurrent}
\end{equation}
This equation can be broken into four terms, each of which can reach values between 0 and 1.
The first term describes how efficiently photons can enter from the feedline into the cavity
before the photons are lost to the environment with the internal loss rate $\kappa_\mathrm{i}$.
Here, $\kappa_\mathrm{c}/\kappa \sim 0.817$ at $f_\mathrm{c}=3.646$~GHz.
The second term, which compares the photon loss rate $\kappa$ of the bare cavity with the effective absorption rate $\kdqd$ of the DQD,
is maximized when these two rates are equal \cite{khan_efficient_2021,wong_quantum_2017}.
With $\kappa=28.1$~MHz and $\kdqd=32.1$~MHz, this term is very close to unity, $\frac{4\kdqd\kappa}{(\kdqd + \kappa)^2} \sim 0.9997$,
indicating that the two systems are well matched.
The third term compares the inter-dot relaxation rate $\gammarelax$ with $\gammalead = \Gamma_\mathrm{Le} + \Gamma_\mathrm{Re}$,
because an excited electron can only result in a dc current if it tunnels out to the reservoir before relaxing back to the ground state.

From the cavity spectroscopy near $\bar{\mu} \sim 0$ presented in Fig.~\ref{fig:fig2}(c), it is not possible to estimate the inter-dot relaxation rate $\gammarelax$,
because cotunneling processes contribute to the extracted decoherence rate $\gammatot/2\pi \sim 829.3~\mathrm{MHz}$ \cite{stockklauser_strong_2017, mi_strong_2017, de_palma_strong_2024}.
Therefore, it is not possible to individually determine $\gammalead$ from the presented measurements.
Instead, we provide a lower bound for $\gammalead$ in the regime where we operate the photodetector. 
In this regime, near the charge triple point ($\bar\mu < 0$), $\gammatot$ can be estimated using the relation $\gammatot/2\pi = 4g^2 / 2\pi \kdqd^- = 1315.5$~MHz.
Assuming that the intrinsic DQD relaxation ($\gammarelax$) and dephasing ($\gammadephasing$) rates remain unchanged at $\bar\mu = 0$ and at the charge triple point,
the observed increase in $\gammatot$ arises from a change in $\gammalead$.
Thus, the difference in total rates sets a lower bound on $\gammalead$, resulting in $\gammalead/2\pi > 486.2~\mathrm{MHz}$ at $\delta=-\deltar$.
Taking into account the spin degeneracy, $\Gamma_\mathrm{g0}/2\pi > 972$~MHz holds accordingly.

The fourth term is the directivity 
\begin{equation}
    \dir = \frac{\Gamma_\mathrm{Re}\Gamma_\mathrm{gL} - \Gamma_\mathrm{Le}\Gamma_\mathrm{gR}}{\gammalead\Gamma_\mathrm{g0}}
\end{equation}
which describes how effectively photo-electrons are tunneling from one reservoir to the other with respect to the opposite direction
\cite{khan_efficient_2021, wong_quantum_2017, haldar_high-efficiency_2024}.
Here, $\Gamma_\mathrm{g0}=\Gamma_\mathrm{gL}+\Gamma_\mathrm{gR}$
represents the total tunneling rate from the reservoirs to the DQD ground state.
Assuming equal QD-reservoir tunneling barriers on the left and right side ($\Gamma_\mathrm{L} = \Gamma_\mathrm{R}$), $\dir$ simplifies to $\deltar/hf_\mathrm{c} \sim 0.877$ at $f_\mathrm{c}=3.646$~GHz
\cite{khan_efficient_2021,haldar_high-efficiency_2024, wong_quantum_2017}.
Further assuming $\frac{\gammalead}{\gammalead + \gammarelax}=1$,
we estimate a total photon detection efficiency of $\eta_\mathrm{calc}\sim 71.3\%$, which is very close to the measured efficiency $\eta^- \sim 67.7\%$,
a value that would be obtained with $\frac{\gammalead}{\gammalead + \gammarelax} \sim 0.955$.
This implies that relaxation of the excited state is dominated by tunneling to the reservoirs ($\gammalead/2\pi>$~486.2~MHz),
while inter-dot relaxation plays only a minor role in determining the photon detection efficiency.
This finding is consistent with prior studies of $\gammarelax$ in similar GaAs DQD devices \cite{scarlino_situ_2022, stockklauser_strong_2017}.

The observed asymmetry in detection efficiency for positive and negative $\delta$
is attributed to unequal tunnel couplings $\Gamma_\mathrm{L}$ and $\Gamma_\mathrm{R}$,
resulting in different values of $\gammalead$ depending on the detuning polarity.
Although such asymmetry generally reduces the directivity $\dir$ and thus $\eta$,
the fact that $\eta \approx \eta_\mathrm{calc}$ suggests that the asymmetry is relatively minor in our device.
Therefore, the dominant factors currently limiting the efficiency from reaching unity
are the coupling of the cavity to the feedline $\kappa_\mathrm{c}$ and the finite directivity $\dir$.
The latter could be improved by either reducing the inter-dot tunnel coupling $t_\mathrm{c}$ or increasing the cavity frequency $f_\mathrm{c}$.
In Appendix~\ref{sec:app_eff}, we demonstrate that detection efficiencies exceeding 90\% are within reach
by engineering devices with lower $t_\mathrm{c}$ and larger $\kappa_\mathrm{c}$, which are feasible in a second generation of devices.

%% file: 1.5_tunability.tex

\section{Frequency Tunability}\label{sec:tunability}

\begin{figure}[!ht]
    \includegraphics[width=\linewidth]{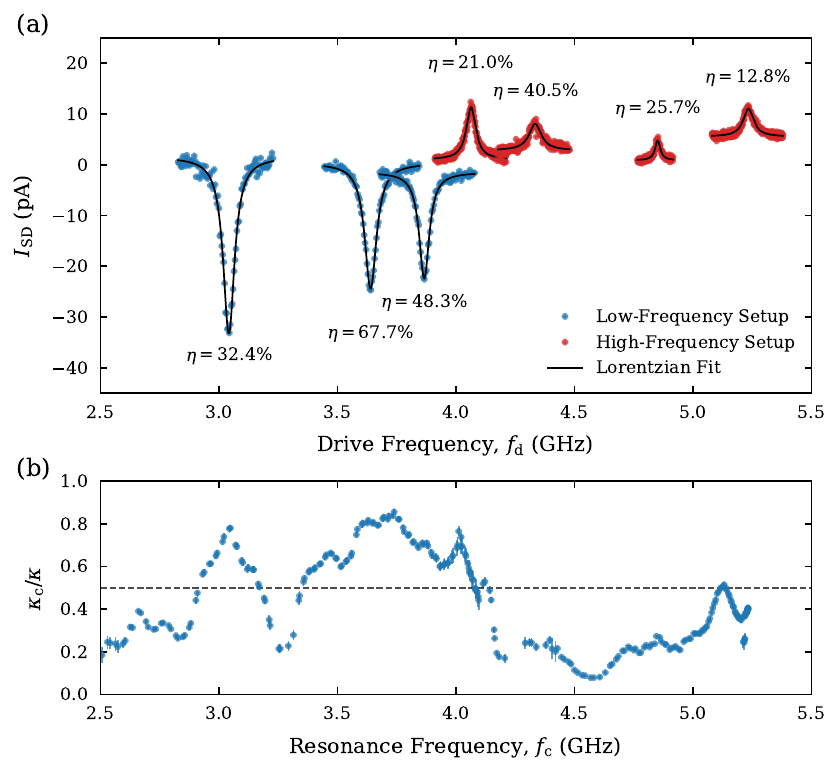}
	\caption{Frequency tunability of the photodetector.
		(a) DQD source-drain current $\isd$ as a function of cavity drive frequency $f_\mathrm{d}$
        measured for different cavity resonance frequencies $f_\mathrm{c}$ and at a constant drive VNA output power.
        The indicated efficiency is obtained by calibrating the input losses for each $f_\mathrm{c}$.
		The blue and red data points are measured in different DQD charge configurations and with different measurement output lines.
		(b) The ratio of the cavity-feedline coupling rate over the total loss rate of the bare cavity, $\kappa_\mathrm{c}/\kappa$, as a function of $f_\mathrm{c}$,
		obtained from fitting the cavity magnetospectroscopy in Fig.~\ref{fig:fig1}(d) (for $f_\mathrm{c}<4.2$~GHz)
        and another similar measurement (for $f_\mathrm{c}>4.2$~GHz).
        The dashed line represents the condition above which the cavity is overcoupled ($\kappa_\mathrm{c} > \kappa_\mathrm{i}$).
	}
    \label{fig:fig4}
\end{figure}

Exploiting the tunability of both the microwave cavity (Fig.~\ref{fig:fig1}(d)) and the DQD charge qubit,
we explicitly demonstrate frequency-selective microwave photon detection.
Figure~\ref{fig:fig4}(a) shows the measured photocurrent as a function of the drive frequency $f_\mathrm{d}$ for different cavity resonance frequencies $f_\mathrm{c}$.
Due to the limited bandwidth of the microwave circulators used in our setup (see Appendix~\ref{sec:app_setup}),
the photocurrent data marked with blue and red dots are acquired in separate cooldowns with different measurement output lines.
To determine the photon detection efficiency at each resonance frequency shown in Fig.~\ref{fig:fig4}(a),
we analyze the power dependence of the PAT current, following the procedure used in Fig.~\ref{fig:fig3}(c).
For all measurements, the drive power at the instrument output is kept constant.
However, due to the presence of spurious modes discussed in Section~\ref{sec:device},
the resulting intra-cavity photon number $n_\mathrm{c}$ is not uniform across different measurements.
This discrepancy explains why the current peak amplitudes do not scale directly with the extracted efficiencies.

As highlighted in Eq.~(\ref{eq:photocurrent}), the ratio between the external coupling rate $\kappa_\mathrm{c}$
and the total cavity decay rate $\kappa$ sets an upper bound on the photon detection efficiency $\eta$ at a given cavity frequency.
Figure~\ref{fig:fig4}(b) displays the extracted values of $\kappa_\mathrm{c}/\kappa$
from a numerical fit of the bare cavity magnetospectroscopy, similar to Fig.~\ref{fig:fig1}(d).
The coupling ratio fluctuates between 0.1 and 0.85,
which we primarily attribute to spurious modes coupling to the cavity that act as additional loss channels,
thereby altering both $\kappa_\mathrm{c}$ and the internal loss rate $\kappa_\mathrm{i}$.
The resonance frequencies used for photoconversion efficiency characterization are selected to maximize $\kappa_\mathrm{c}/\kappa$.
The extracted cavity loss rates $\kappa_\mathrm{c}$, $\kappa_\mathrm{i}$,
and $\kappa$ at various frequencies are summarized in Appendix~\ref{sec:app_kappas}.

%% file: 1.6.1_discussion.tex
\section{Discussion}\label{sec:discussion}

As we show in Appendix~\ref{sec:app_eff}, the photon detection efficiency $\eta$ is sensitive to changes in several system parameters (see Eq.~(\ref{eq:photocurrent})),
most notably $g_0$, the cavity loss rates $\kappa_\mathrm{c}$ and $\kappa_\mathrm{i}$, and the inter-dot relaxation rate $\gamma_-$.
A large $g_0$ ensures that the effective coupling strength at finite DQD detuning
($g = g_0\cdot 2t_\mathrm{c}/\sqrt{\delta^2 + 4t_\mathrm{c}^2}$) remains large across a wide range of detuning values.
This enables efficient interaction between the cavity and the DQD, allowing one to achieve a condition where the effective photon absorption rate $\kdqd=4g^2/\gammatot$
can be matched to the total cavity linewidth $\kappa$ with a large $\gammalead$.
In our device, this condition is achieved thanks to the high tunability of $\gammalead$, and carefully engineered tunnel barriers, enabled by the gate-defined architecture .
A large $\gammalead$ also makes the detection efficiency robust against inter-dot relaxation,
as the relaxation of the excited state is dominated by tunneling to the reservoirs rather than by internal processes.

An additional term influencing the efficiency is the directivity $D \sim \deltar / h f_\mathrm{c}$,
which compares how efficiently a photo-excited electron tunnels in one direction with respect to the other.
Large detuning $\deltar$ and low inter-dot tunnel coupling $t_\mathrm{c}$ are required to maximize $\dir$,
however, reducing $t_\mathrm{c}$ also lowers the coupling $g$.
Thus, a large intrinsic $g_0$ is advantageous for supporting strong interaction even at reduced $t_\mathrm{c}$,
enabling high directivity without compromising coupling strength.

The main limitation to the efficiency of the studied DQD-cavity photon detector stems from the ratio $\kappa_\mathrm{c} / \kappa$,
which quantifies how efficiently photons enter the cavity from the feedline.
This ratio is ultimately limited by losses intrinsic to the cavity.
However, as illustrated in Fig.~\ref{fig:suppfig_efficiency}(a), increasing $\kappa_\mathrm{c} / \kappa$
through modest geometric optimization of the feedline coupling would immediately push $\eta$ beyond 80\%,
assuming the other parameters remain unchanged.
Moreover, further reducing the charge qubit relaxation rate, e.g.,
by increasing the mutual capacitance between the dots~\cite{scarlino_situ_2022},
can help approach near-unity detection efficiency, as confirmed in Fig.~\ref{fig:suppfig_efficiency}(b).

Overall, our analysis shows that efficiencies exceeding 90\% are realistically achievable
in this architecture with straightforward device improvements:
enhanced cavity-feedline coupling, a larger cavity resonance frequency and a reduced qubit decoherence rate.
In addition, our architecture is inherently frequency-selective, not susceptible to quasiparticle poisoning
\cite{stanisavljevic_efficient_2024, pankratov_detection_2025},
and supports broadband tunability over a range of $3-5.2$~GHz.

Two key figures of merit of a photodetector are the dead time and the noise equivalent power (NEP).
The dead time $\tau_\mathrm{dead}$ corresponds to the time required for the excited DQD system to reset to its ground state,
which is given by $\tau_\mathrm{dead} = 2\pi(1/\Gamma_\mathrm{0e} + 1/\Gamma_\mathrm{g0})$.
In our device, $\Gamma_\mathrm{0e}/2\pi$ ($\Gamma_\mathrm{g0}/2\pi$) is expected to be higher than 486~MHz (972~MHz),
implying that $\tau_\mathrm{dead} < 3~\mathrm{ns}$. 
This short dead time enables nearly continuous operation of the detector, eliminating the need for heralding,
pulsed gating, or explicit reset procedures~\cite{besse_single-shot_2018, balembois_cyclically_2024, inomata_single_2016}.
The NEP represents the minimum input power required to achieve a signal-to-noise ratio of one for an integration time of 1~s.
In our setup, where the current through the DQD is continuously monitored, the noise is proportional to the current fluctuations $\delta \isd$ in the dc measurement chain, where we observe $\delta \isd \sim 50$~fA when measured with a measurement bandwidth of $B = 5$~Hz.
With a current responsivity $R = e\eta/hf_\mathrm{c} \sim 45$~kA/W at $f_\mathrm{c} = 3.646$~GHz  $\mathrm{NEP} = \delta \isd/R\sqrt{B} \sim 5 \times 10^{-19}~\mathrm{W}/\sqrt{\mathrm{Hz}}$,
consistent with values reported for other platforms \cite{stanisavljevic_efficient_2024, haldar_high-efficiency_2024}.

\section{Conclusions}

In this work, we demonstrate high-efficiency microwave photon detection in the single-photon regime
using a hybrid semiconductor-superconductor circuit QED device, achieving detection efficiencies up to $\eta \sim 70\%$.
This device substantially outperforms previously reported DQD-based detectors \cite{khan_efficient_2021, haldar_high-efficiency_2024}
and sets a new benchmark for the technology.
This improvement is enabled by the combination of a large charge-photon coupling strength $g_0$ obtained due to the high-impedance cavity,
broad tunability of both the cavity frequency and the DQD tunneling rates, as well as the low DQD charge relaxation rate.
These features are critical for simultaneously optimizing all the processes affecting the photon detection efficiency,
as expressed in Eq.~(\ref{eq:photocurrent}).

The high detection efficiency achieved with a semiconductor DQD in this work opens new possibilities in the realm of quantum technology. 
Our detector, approaching unity efficiency, offers a practical path for measuring correlations between microwave photons---significantly reducing the experimental overhead \cite{hoi_generation_2012}
without the need for parametric amplification chains \cite{balembois_cyclically_2024}. 
Further architectural improvements, such as embedding a high-bandwidth charge sensor~\cite{haldar_continuous_2024, havir_near-unity_2025, geng_high-fidelity_2025},
will enable single-shot detection of itinerant microwave photons and facilitate integration into larger quantum photonic networks. 
Notably, since DQD microwave photon detectors can be built on the same substrate as QD spin qubit architectures, interfacing these detectors with spin qubits could unlock new avenues
for microwave photonics---domains that have so far been largely limited to superconducting qubit systems \cite{gu_microwave_2017, opremcak_measurement_2018}---and for QD spin qubit operations \cite{opremcak_measurement_2018, wang_single-electron_2023}. 
Our work thus establishes a clear and scalable route to robust, high-performance microwave photon detection, with immediate applications in quantum communication, quantum sensing, and microwave quantum optics.

\section*{Author's Contributions}

F.O. and W.J. performed the measurements, analyzed the data and contributed equally to this work.
A.T., F.D. and F.O. developed the fabrication processes.
A.T. and F.O. fabricated the device.
C.R. and W.W. designed the GaAs/AlGaAs heterostructure and performed the growth.
F.O., W.J. and P.S. conceived the experiments and wrote the manuscript with inputs from all authors.
P.S., V.F.M. and D.Z. supervised the work.

\section*{Acknowledgments}

We thank
Pan Zhang, Gianluca Rastelli, Martin Leijnse and Alberto Mercurio
for fruitful discussions.
P.S. acknowledges support from the Swiss State Secretariat for Education, Research and Innovation (SERI) under contract number MB22.00081~/~REF-1131-52105. 
P.S and D.Z. acknowledge support from the NCCR SPIN, a National Centre of Competence in Research, funded by the Swiss National Science Foundation (SNSF) with grant number 225153.
P.S. also acknowledges support from the SNSF through the grants Ref. No. 200021\_200418~/~1 and Ref. No. 206021\_205335~/~1.
W.J. acknowledges support from EPFL QSE Postdoctoral Fellowship Grant.
D.Z. and A.T. acknowledge support from the SNSF through grants Ref. No. 215757.
V.F.M thanks NanoLund for financial support.


%% file: 2_appendix.tex

\section{Device Fabrication}
\label{sec:app_fab}

The device is fabricated on a GaAs/Al$_x$Ga$_{1-x}$As heterostructure with a 2-dimensional electron gas (2DEG) at the GaAs/Al$_x$Ga$_{1-x}$As interface, buried 90~nm below the surface.
195/53~nm AuGe/Pt ohmic contacts are patterned by electron-beam lithography (EBL), followed by electron-beam evaporation and lift-off.
The device is then annealed in forming gas (5\% H$_2$, 95\% N$_2$) at 490°C for 1~min to let the metal diffuse into the heterostructure and ensure good contact with the 2DEG.
The self-accumulated 2DEG is etched away with a Piranha solution of H$_2$SO$_4$ 30\%~:~H$_2$O$_2$~:~DI (1:8:640) leaving a only a small mesa region that forms a well-defined conductive channel for the DQD.
The single-layer gates are patterned in two steps by EBL, evaporation and lift-off.
This ensures that the thin 2/26~nm Ti/Au gates are patched on the mesa step ($\sim 110$~s) by a second 3/110~nm layer, that is routed out to the bonding pads.
A superconducting Al layer of 120~nm is deposited using EBL, evaporation and lift-off, to form the ground plane and the feedline for the cavity in a single step.
Lastly, the JJ array cavity is fabricated using the conventional Dolan-bridge double angle evaporation method.
Two Al layers of 35~nm and 130~nm, respectively, are deposited at an angle of 45° with an oxidation step in between to form the tunneling barrier.
This tunneling oxide is grown by filling the chamber with O$_2$ at a pressure of 2~Torr for 20~min (static oxidation) without breaking the vacuum.
Finally, after the second Al deposition, the device is once again exposed to 10~Torr of O$_2$ for 10~min to form a clean oxide layer at the top interface.

\section{Input-Output Model}\label{sec:app_io}

To model the cavity response in our hybrid system, we employ an input-output model based on a quantum Rabi model to also consider the counter-rotating term,
which is non-negligible in a strongly coupled system \cite{frisk_kockum_ultrastrong_2019,forn-diaz_observation_2010,niemczyk_circuit_2010}.
We first show the Hamiltonian which describes a general cavity-qubit hybrid system.
Assuming the transverse coupling between the photon and the qubit, the Hamiltonian can be written as Eq.~(\ref{eq:full_hamiltonian}) \cite{blais_circuit_2021}
\begin{equation}
\label{eq:full_hamiltonian}
    H_\mathrm{full} /\hbar = \omega_\mathrm{c}a^\dagger a+\frac{\omega_\mathrm{q}}{2}\sigma_z+g(a^\dagger + a)\sigma_x,
\end{equation}
where, $\omega_\mathrm{c}=2\pi f_\mathrm{c}$ and $\omega_\mathrm{q}=2\pi f_\mathrm{q}$ are the cavity and qubit frequencies, respectively, $a$ ($a^\dagger$) is the photon annihilation (creation) operator
and $\sigma_z$ and $\sigma_x$ are Pauli operators describing the qubit state.
The effective charge-photon coupling strength $g=g_0\sin\theta$ is normalized by the mixing angle $\theta$,
where $\sin\theta = 2t_\mathrm{c}/\sqrt{\delta^2+4 t_\mathrm{c}^2}$ and $g_0$ is the charge-photon coupling strength at $\delta=0$~\cite{stockklauser_strong_2017}.
Applying the rotating wave approximation (RWA) in the frame rotating with the drive frequency $\omega_\mathrm{d}=2\pi f_\mathrm{d}$ yields the typical Jaynes-Cummings (JC) Hamiltonian depicted in Eq.~(\ref{eq:JC_hamiltonian}),
with the qubit lowering (raising) operator $\sigma_-$ ($\sigma_+$). 
Note that the RWA eliminates the counter-rotating terms proportional to $a\sigma_-$ and $a^\dagger\sigma_+$.
\begin{equation}
\label{eq:JC_hamiltonian}
    H_\mathrm{JC}/\hbar=\Delta_\mathrm{c}a^\dagger a+\frac{\Delta_\mathrm{q}}{2}\sigma_z+g(a^\dagger \sigma_- + a\sigma_+),
\end{equation}
where $\Delta_\mathrm{c} = \omega_\mathrm{d} - \omega_\mathrm{c}$ and $\Delta_\mathrm{q} = \omega_\mathrm{d} - \omega_\mathrm{q}$.
We solve Eq.~(\ref{eq:full_hamiltonian}) and Eq.~(\ref{eq:JC_hamiltonian}) to obtain the eigenspectrum shown in full and dashed lines respectively in Figs.~\ref{fig:fig2}(c) and (d),
using the parameters reported in Table~\ref{tab:app_parameters}.

In general, the input-output relation that describes a reflection-type cavity coupled to a qubit is written as
\begin{equation}
\label{eq:io}
    S_{11} =\frac{\Delta_\mathrm{c} + i(\kappa_\mathrm{c}-\kappa_\mathrm{i})/2 + g \chi} {\Delta_\mathrm{c} + i\kappa/2 + g \chi},
\end{equation}
where $\chi$ is the susceptibility of the qubit to the cavity photon, $\kappa_\mathrm{c}$ is the cavity-feedline coupling rate,
$\kappa_\mathrm{i}$ is the internal cavity loss and $\kappa=\kappa_\mathrm{c}+\kappa_\mathrm{i}$ is the total cavity loss rate.
The two Hamiltonians of Eq.~(\ref{eq:JC_hamiltonian}) and Eq.~(\ref{eq:full_hamiltonian})
yield the distinct susceptibilities Eq.~(\ref{eq:chi_JC}) \cite{de_palma_strong_2024} and Eq.~(\ref{eq:chi_full}) \cite{kohler_dispersive_2018}, respectively:
\begin{align}
    \label{eq:chi_JC}
    \chi_\mathrm{JC} &= \frac{g}{-\Delta_\mathrm{q} - i\gammatot} \\
    \label{eq:chi_full}
    \chi_\mathrm{full} &= \frac{g}{-\Delta_\mathrm{q} - i\gammatot} + \frac{g}{-(\omega_\mathrm{d} + \omega_\mathrm{q}) - i\gammatot},
\end{align}
where $\gammatot$ represents the decoherence rate of the qubit \cite{de_palma_strong_2024}.
For a hybrid system with strong coupling between the subsystems, i.e. with $g/\omega_\mathrm{c}$ approaching 0.1 or higher,
the counter-rotating terms become non-negligible, and the JC model fails to describe the system \cite{frisk_kockum_ultrastrong_2019, forn-diaz_observation_2010, niemczyk_circuit_2010}.

\begin{figure}[!ht]
    \includegraphics{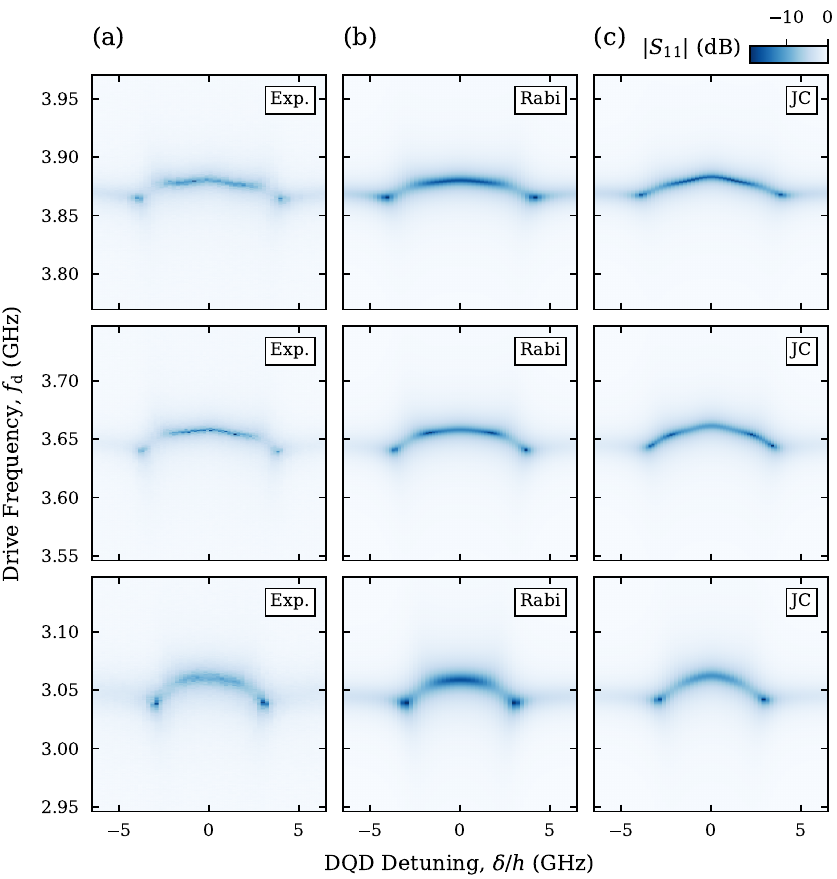}
	\caption{(a) Measured normalized cavity reflectance $|S_{11}|$ as a function of the cavity drive frequency $f_\mathrm{d}$ and DQD detuning $\delta$
    for different cavity resonance frequencies $f_\mathrm{c} = 3.878,~ 3.646,$ and $3.032$~GHz.
    Simulated $|S_{11}|$ for each $f_\mathrm{c}$ using an input-output model built with (b) full-Rabi Hamiltonian taking into account the counter-rotating terms and (c) Jaynes-Cummings Hamiltonian neglecting the counter-rotating terms.
	}
    \label{fig:suppfig_spectro}
\end{figure}

This can be observed in our system, where $g/\omega_\mathrm{q}\sim 0.1-0.12$ at $\delta = 0$.
In Fig.~\ref{fig:suppfig_spectro}(a), we show a single tone spectroscopy measurement of the DQD configuration shown in Fig.~\ref{fig:fig2}
for three different cavity resonance frequencies $f_\mathrm{c} = $ 3.032, 3.646 and 3.872~GHz.
We numerically fit the input-output model Eq.~(\ref{eq:io}) to the spectra with $\chi=\chi_\mathrm{full}$ to yield the simulated spectra in Fig.~\ref{fig:suppfig_spectro}(b).
We also show the cavity spectra simulated by using the input-output model with $\chi=\chi_\mathrm{JC}$ in Fig.~\ref{fig:suppfig_spectro}(c),
which deviates from the experimental data in Fig.~\ref{fig:suppfig_spectro}(a),
not properly capturing the Bloch-Siegert shift around $\delta=0$ \cite{forn-diaz_observation_2010}.
The estimated device parameters of the photodetector shown in Fig.~\ref{fig:fig2} and Fig.~\ref{fig:fig3} are reported in Table~\ref{tab:app_parameters}.
$g_0$, $t_\mathrm{c}$ and $\gammatot$ have been estimated using the full input-ouput model with $\chi_\mathrm{full}$.

\begin{table}[!ht]
    \begin{tabular}{rrrr}
        \toprule
 $f_\mathrm{c}$      & 3032~MHz            & 3646~MHz            & 3872~MHz           \\
        \midrule
 $\kappa/2\pi$            & 28.6 $\pm$ 0.3  & 28.1 $\pm$ 0.3  & 25.1 $\pm$ 0.3  \\
 $\kappa_\mathrm{c}/2\pi$ & 22.8 $\pm$ 0.2  & 23.0 $\pm$ 0.2  & 18.3 $\pm$ 0.1  \\
 $\kdqd^+/2\pi$           & 28.1 $\pm$ 2.5  & 27.3 $\pm$ 1.5  & 24.1 $\pm$ 1.2  \\
 $\kdqd^-/2\pi$           & 29.3 $\pm$ 1.7  & 32.1 $\pm$ 1.3  & 26.7 $\pm$ 1.2  \\
 $g_0/2\pi$               & 183.7 $\pm$ 0.5 & 213.7 $\pm$ 0.3 & 210.6 $\pm$ 0.5 \\
 $t_\mathrm{c}/h$         & 878.0 $\pm$ 4.4 & 878.0 $\pm$ 4.4 & 878.0 $\pm$ 4.4 \\
 $\gammatot/2\pi$         & 722.2 $\pm$ 5.4 & 829.3 $\pm$ 3.6 & 893.0 $\pm$ 6.3 \\
 $\eta^+$ (\%)            & 22.1 $\pm$ 1.9  & 55.8 $\pm$ 4.0  & 37.0 $\pm$ 2.5  \\
 $\eta^-$ (\%)            & 26.9 $\pm$ 2.4  & 67.7 $\pm$ 4.8  & 44.3 $\pm$ 3.0  \\
        \bottomrule
    \end{tabular}
    \caption{Device parameters from fits at three different cavity resonance frequencies $f_\mathrm{c}$.
    $\kappa$ and $\kappa_\mathrm{c}$ are extracted from bare cavity spectroscopy measurements,
    while $g_0$, $t_\mathrm{c}$ and $\gammatot$ are obtained from the fits reported in Fig.~\ref{fig:suppfig_spectro}(b) using an input-output model that retains the counter-rotating terms.
    $\kdqd$ and $\eta$ are extracted from the fits reported in Figs.~\ref{fig:fig3}(e) and (b), respectively, for $\delta=\pm\deltar$.
    All values except for $\eta$ are reported in MHz.
    }
    \label{tab:app_parameters}
\end{table}

\section{Measurement Setup}\label{sec:app_setup}

\begin{figure}[!ht]
    \includegraphics{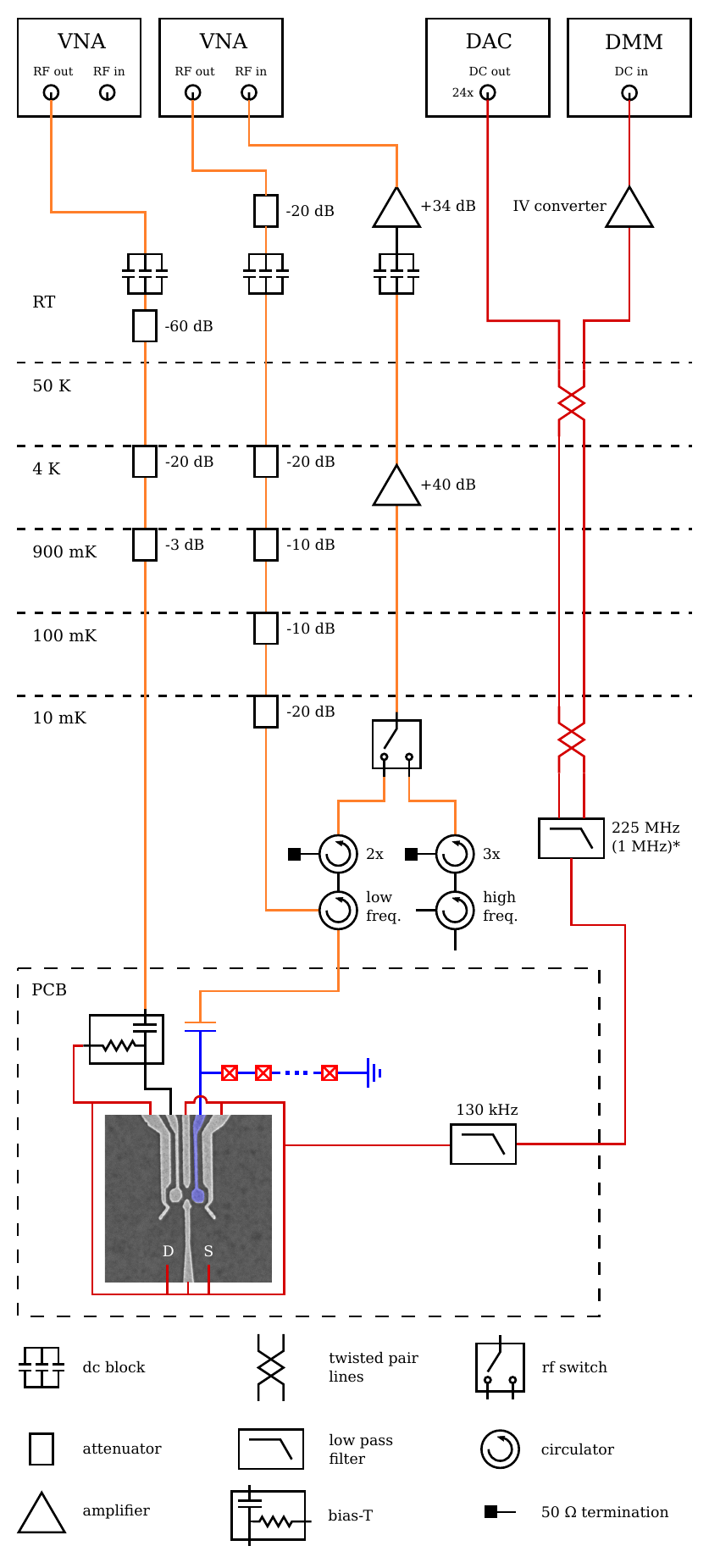}
	\caption{Schematic of the cryogenic measurement setup. The orange lines denote the coaxial cables hosting rf signals,
            while the red lines denote the twisted pair dc cables.
	}
    \label{fig:suppfig_setup}
\end{figure}

The measurements reported in this work are performed in a bottom loading dilution refrigerator (Bluefors LD250) at base temperatures around 10 mK (see Fig. ~\ref{fig:suppfig_setup}).
The device is mounted on a printed circuit board (PCB, QDevil QBoard II), which consists of a daughterboard hosting the device and a motherboard,
to which the daughterboard is connected via spring contacts.
The motherboard features an RC low pass filtering stage (130~kHz cut-off) as well as bias-Ts for combining rf and dc signals.
The PCB is mounted inside a fast sample exchange (FSE) probe ($T\sim50$~mK), which is inserted into the cryostat without warming up the full system.

Gate and bias voltages for the QDs are generated by a 24-channel digital-to-analog converter (DAC, QDevil QDAC-II)
and are passed to the PCB via twisted pair cables made of phosphor bronze and an additioanal LC low-pass filter stage at the mixing chamber plate at $T\sim 10$~mK.
For the dc connections of the DQD gate lines, two different kinds of LC filters are used,
a QDevil filter box (225~MHz cut-off) and a MFT25-150$\Omega$ box (1~MHz cut-off) by Basel Precision Instruments.
The drain current through the device is measured by a digital multimeter (DMM, Keysight 34465A)
after being amplified by an IV converter (Basel Precision Instruments SP~983C).
The in-plane magnetic field is applied by a 6-1-1~T vector magnet by American Magnetics.

Spectroscopy measurements are performed with a vector network analyzer (VNA, Rohde \& Schwarz ZNB20).
The VNA output is attenuated at room temperature, followed by a dc block (Inmet 8039, inner-outer, 10~MHz cut-off), followed by another attenuation chain in the cryostat (-60~dB).
The dc block is essential to avoid ground loops that would introduce low-frequency noise in the DQD gate lines or ohmic contacts.
To measure the signal reflected by the cavity in a wide frequency range, we use two different sets of circulators,
a triple junction low-frequency circulator (Quinstar QCY-G0250403AM, $2.5-4$~GHz)
and two double junction high-frequency circulators (Low Noise Factory CICIC4\_12A, $4-12$~GHz).
A microwave switch (Radiall R570F32000) allows to choose which set of circulators to connect
to the broadband HEMT amplifier at 4~K (Low Noise Factory LNC0.3\_14B, $0.3-14$~GHz, +40~dB).
After amplification at 4~K, the reflected signal passes through another dc block (Inmet 8039)
and a low noise amplifier at room temperature (Agile AMT\_A0253, $0.1-20$~GHz, +34~dB).
For two-tone spectroscopy measurements, the second tone is applied by another VNA (Rohde \& Schwarz ZNB20) in a fixed frequency mode.
The DQD drive line, with a total cryogenic attenuation of 23~dB,
is connected to the left plunger gate ($V_\mathrm{pL}$ in Fig.~\ref{fig:fig1}(b)) of the DQD via bias-T on the PCB.
Contrary to the cavity drive line, for the qubit drive line the attenuators at room temperature are placed in between dc block and the cryostat to attenuate standing waves between the dc block and the bias-T.

\section{Noise-Induced Currents in Magnetic Field}\label{sec:app_noise}

\begin{figure}[!ht]
    \includegraphics{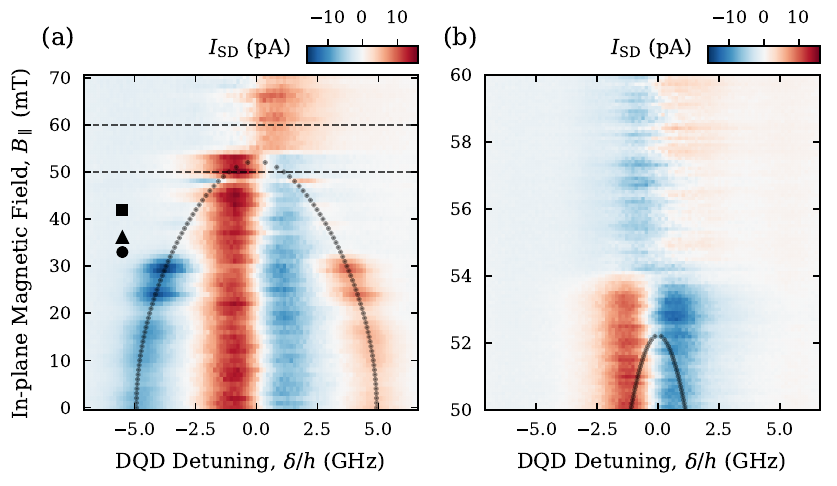}
	\caption{(a) DQD source-drain current $\isd$ as a function of the in-plane magnetic field $B_\parallel$ parallel to the JJ array (see Fig.~\ref{fig:fig1}(c)) and the DQD detuning $\delta$ in the same region as in Fig.~\ref{fig:fig3}, without any cavity drive. 
            The grey dots denote the magnetic field for which qubit and cavity are in resonance.
            Black square, triangle and circle illustrate the three different $B_\parallel$ at which we extensively investigated the photon detection,
            also reported in Fig.~\ref{fig:suppfig_spectro}.
            (b) Zoom-in of the region enclosed by the horizontal dashed lines in (a).
	}
    \label{fig:suppfig_noise}
\end{figure}

As evident from the charge stability diagram shown in Fig.~\ref{fig:fig2}(a), a finite dark current is present near $\delta \sim 0$,
even with negligibly small excitation in the cavity ($n_\mathrm{c} \ll 1$). 
The dark current shows an antisymmetric behavior as a function of $\delta$, which indicates the presence of detuning-dependent transport processes enabled by 
by phonons \cite{dorsch_heat_2021} or electrical noise \cite{entin-wohlman_heat_2017} in the setup.
While pinpointing the exact origin of these dark currents is out of the scope of this work,
we show that they are enhanced by the superconducting cavity by measuring $\isd$ around $\delta=0$ as a function of the in-plane magnetic field $B_\parallel$.
Figure~\ref{fig:suppfig_noise}(a) shows the same current around $\delta=0$ as Fig.~\ref{fig:fig2}(a),
but as we increase the magnetic field the nature of the current changes.
A zoom-in at higher magnetic fields (see Fig.~\ref{fig:suppfig_noise}(b)) reveals that this current is reduced and even inverted in polarity when reaching about 54~mT,
which is the typical in-plane critical field for similar JJ array devices.
One possible explanation is that the noise originates from the drive line connected to the left plunger gate (see Appendix.~\ref{sec:app_setup}),
which is poorly thermalized.
The noise can couple to the cavity, which has a stronger impact on the DQD device due to its high impedance.
When the JJ array turns resistive, the noise in the cavity is mostly dissipated and only the noise in the drive line, i.e. the left plunger gate, remains.

Another evidence for the noise in the cavity affecting the device is visible in Fig.~\ref{fig:suppfig_noise}(a).
On top of the current around $\delta=0$, non-zero current also appears at $\delta=\pm\deltar$,
where the qubit frequency matches the cavity resonance frequency (indicated by grey dots in Figs.~\ref{fig:suppfig_noise}(a) and (b)),
in spite of the absence of a cavity drive tone.
This means that at these frequencies, the cavity is in fact populated ($n_\mathrm{c}>0$), potentially due thermal photons,
resulting in a clearly visible PAT current, due to the high photon detection efficiency.
Interestingly, the visibility of these PAT currents varies strongly depending on $f_\mathrm{c}$,
suggesting that the extra cavity photon population is mediated by the aforementioned spurious modes present in the environment (see Fig.~\ref{fig:fig1}(d)).
We stress, that the characterization of the photodetector in the high-efficiency configuration reported in Fig.~\ref{fig:fig3} and Table~\ref{tab:app_parameters}
is performed at frequencies for which this noise-induced current is suppressed (indicated by black square, triangle and circle in Fig.~\ref{fig:suppfig_noise}(a)).
This explains the comparably low dark current observed in the reported measurements.

\section{Alternative DQD Configuration}\label{sec:app_other_inter-dot}

\begin{figure}[!ht]
    \includegraphics{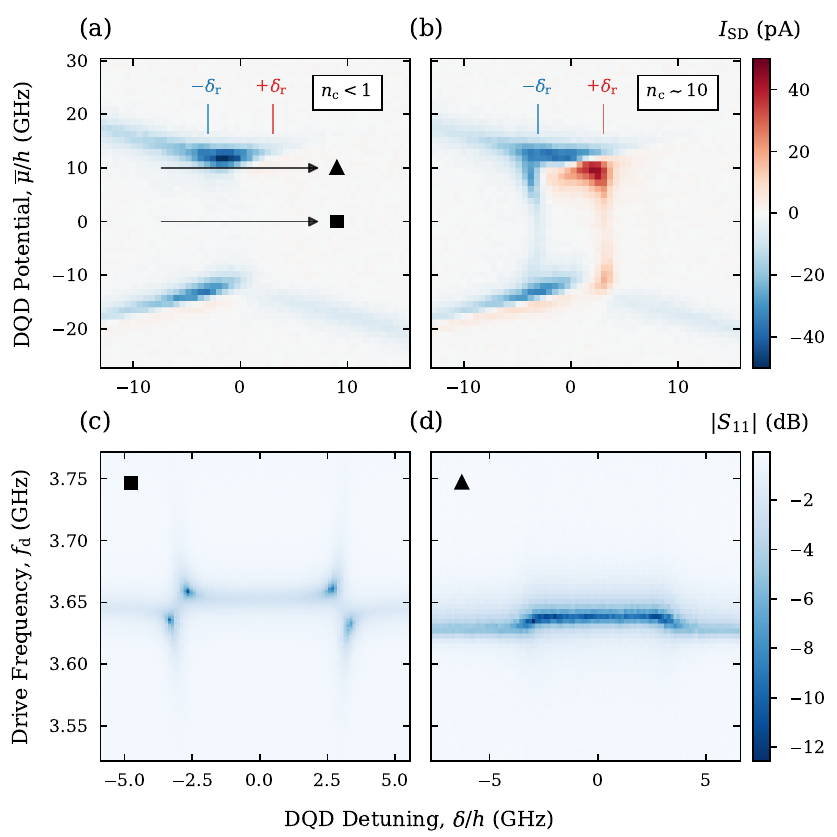}
	\caption{Alternative DQD configuration to the one shown in Fig.~\ref{fig:fig2} with reduced cotunneling current.
            DQD charge stability diagrams measured as a function of the DQD detuning $\delta$ and potential $\bar{\mu}$ while monitoring the DQD source-drain current $\isd$ with (a) low ($n_\mathrm{c} < 1$) and (b) high ($n_\mathrm{c} \sim 10$) cavity drive power.
            $\pm \deltar$ denote the DQD detuning values at which the charge qubit energy matches that of the cavity photon.
            In panel (b), clear features related to photon-assisted tunneling appear at $\pm \deltar$.
            (c-d) Normalized cavity reflectance $|S_{11}|$ measured as a function of the drive frequency $f_\mathrm{d}$ while sweeping $\delta$ along the arrow denoted by (c) square ($\bar \mu /h\sim 0$~GHz) and (d) triangle ($\bar \mu /h \sim 10$~GHz) in panel (a). 
	}
    \label{fig:suppfig_other_inter-dot}
\end{figure}

A different DQD configuration with respect to the one presented in Fig.~\ref{fig:fig2}
has been studied and is presented in Fig.~\ref{fig:suppfig_other_inter-dot}.
This configuration does not feature any antisymmetric current around $\delta=0$ at low cavity drive power,
in contrast to the previous DQD configuration (see Fig.~\ref{fig:suppfig_other_inter-dot}(a)).
At higher drive power ($n_\mathrm{c} \sim 10$), the stability diagram exhibits a strong PAT current symmetrically around the charge triple points (see Fig.~\ref{fig:suppfig_other_inter-dot}(b)).
In contrast to the configuration in Fig.~\ref{fig:fig2}, this PAT current is significantly reduced when going toward the center of the inter-dot transition ($\bar{\mu} = 0$),
which means that cotunneling effects are less prominent in this configuration.
This observation is also reflected in the spectroscopy measurements taken along the black arrows in Fig.~\ref{fig:suppfig_other_inter-dot}(a).
Close to the charge triple point (Fig.~\ref{fig:suppfig_other_inter-dot}(d), triangle cut from Fig.~\ref{fig:suppfig_other_inter-dot}(a)),
the charge qubit decoherence rate is significantly larger
than the one at $\bar{\mu} = 0$ (Fig.~\ref{fig:suppfig_other_inter-dot}(c), square cut from Fig.~\ref{fig:suppfig_other_inter-dot}(a)),
due to the large relaxation rate to the reservoirs.
The different behavior compared to Fig.~\ref{fig:fig2} likely stems from a combination of smaller tunneling rate to the reservoirs
and larger inter-dot charging energy $U$, which reduces cotunneling effects \cite{amasha_pseudospin-resolved_2013}.

\section{ac Stark Shift Measurements}\label{sec:app_ac-stark}

\begin{figure}[!ht]
    \includegraphics{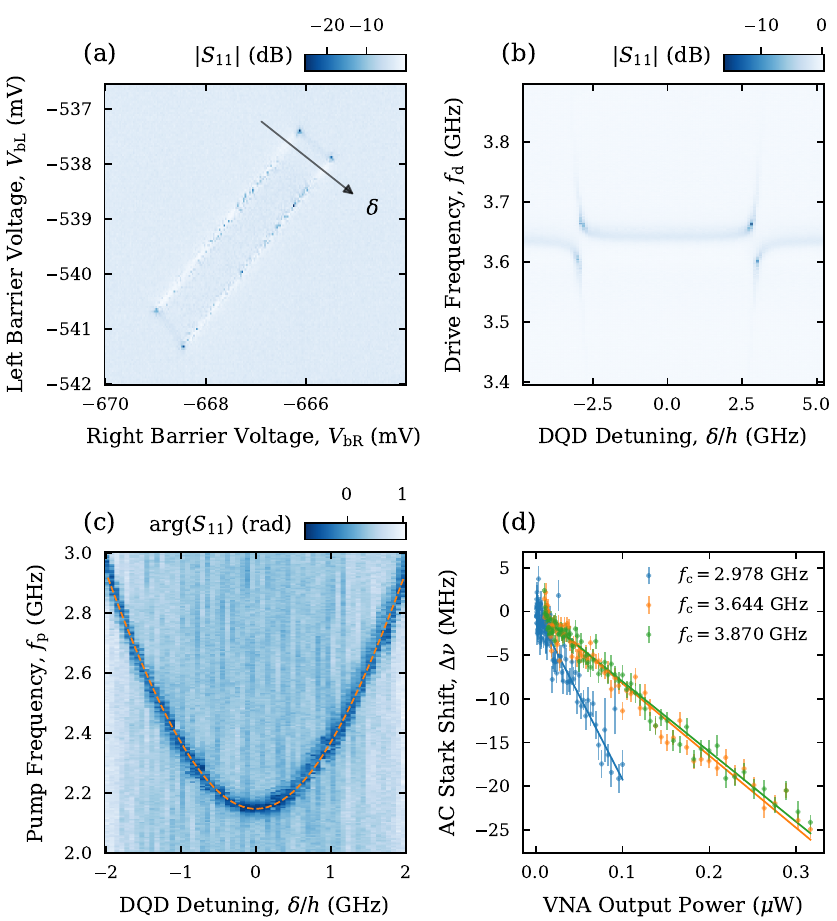}
	\caption{ac Stark shift measurements for input loss characterization.
            (a) DQD charge stability diagram of a new configuration in the same device,
            measured as a function of the right and left barrier gate voltage $V_\mathrm{bR}$ and $V_\mathrm{bL}$ (see Fig.~\ref{fig:fig1}(b))
            by monitoring the normalized cavity reflectance $|S_{11}|$.
            (b) $|S_{11}|$ as a function of the drive frequency $f_\mathrm{d}$ and DQD detuning $\delta$ tuned along the black arrow in panel (a).
            (c) DQD charge qubit excitation frequency measured by monitoring the phase of the cavity reflectance arg($S_{11}$)
            as a function of $\delta$ and the qubit pump frequency $f_\mathrm{p}$.
            The cavity drive tone is kept at $f_\mathrm{d} = f_\mathrm{c} \sim 3.644$~GHz.
            The dashed orange curve represents a numerical fit to the charge qubit frequency $f_\mathrm{q} = \sqrt{\delta^2 + 4t_\mathrm{c}^2}/h$,
            from which we extract $2t_\mathrm{c}/h \sim 2.2$~GHz.
            (d) ac Stark shift $\Delta\nu$ of the qubit measured by recording $f_\mathrm{q}$ at $\delta=0$ as a function of the VNA output power
            applied to the cavity at $f_\mathrm{d} = f_\mathrm{c} \sim$ 2.978~GHz (blue dots), 3.644~GHz (orange dots) and 3.870~GHz (green dots).
            The solid lines represent linear fits to Eq.~(\ref{eq:ac_stark}) from which we extract the input loss factors $\beta$ for the frequencies in the legend.
	}
    \label{fig:suppfig_ac-stark}
\end{figure}

In the dispersive regime, i.e. when the detuning between qubit and cavity $\Delta_\mathrm{qc}= \omega_\mathrm{q}-\omega_\mathrm{c}$ is large enough, such that $\Delta_\mathrm{qc} \gg g$,
the Jaynes-Cummings Hamiltonian in Eq.~(\ref{eq:JC_hamiltonian}) can be reduced to
\begin{equation}
    H = \hbar\omega_\mathrm{c}a^\dagger a + \frac{1}{2}\hbar(\omega_\mathrm{q} + \chi + 2\chi a^\dagger a)\sigma_z,
    \label{eq:Ham_dispersive}
\end{equation}
where $\chi = g^2/\Delta_\mathrm{qc}$ is the dispersive shift \cite{schuster_ac_2005}.
We can thus write the dressed qubit frequency as $\tilde{\omega}_\mathrm{q} = \omega_\mathrm{q} + \chi + 2n_\mathrm{c}\chi$,
with the average cavity photon number $n_\mathrm{c}=\langle a^\dagger a\rangle$.
The coupling to the cavity therefore leads to a power-dependent ac Stark shift of
\begin{equation}
    \Delta\nu = (\tilde{\omega}_\mathrm{q} - \tilde{\omega}_\mathrm{q,0})/2\pi = 2n_\mathrm{c} g^2/(2\pi\Delta_\mathrm{qc}),
    \label{eq:ac_stark}
\end{equation}
where $\tilde{\omega}_\mathrm{q,0}$ is the measured qubit frequency at low cavity drive power ($n_\mathrm{c} \ll 1$).
The average photon number can be related to the input power $P_\mathrm{d}$ using
\begin{equation}
    n_\mathrm{c} = \frac{4\kappa_\mathrm{c}}{\kappa^2}\dot{N} = \frac{4\kappa_\mathrm{c}}{\kappa^2} \frac{P_\mathrm{d}}{hf_\mathrm{c}}
     = \frac{4\kappa_\mathrm{c}}{\kappa^2} \frac{\beta P_\mathrm{VNA}}{hf_\mathrm{c}},
     \label{eq:nc}
\end{equation}
where $P_\mathrm{VNA}$ is the VNA output power and $\beta=P_\mathrm{d}/P_\mathrm{VNA}$ is the total attenuation factor
taking into account all the losses between the VNA and the cavity feedline.
Therefore, measuring the ac Stark shift in a stable DQD configuration is a useful tool for estimating the effective power
reaching the device at a specific cavity frequency \cite{schuster_ac_2005}.

In a separate cooldown from the one in Fig.~\ref{fig:fig2} and Fig.~\ref{fig:fig3},
we tune the device to a more coherent DQD charge configuration (shown in Fig.~\ref{fig:suppfig_ac-stark}(a))
obtained by defining large mutual capacitance between the two QDs
to reduce its sensitivity to charge noise \cite{scarlino_situ_2022} and drastically reducing the tunneling rate to the QD reservoirs.
In Fig.~\ref{fig:suppfig_ac-stark}(b), we present a single tone spectroscopy measurement, implemented as in Fig.~\ref{fig:fig2}(c),
as a function of $\delta$ measured along the black arrow in Fig.~\ref{fig:suppfig_ac-stark}(a).
A clear vacuum-Rabi splitting at finite $\delta$ indicates that the qubit frequency lies well below $f_\mathrm{c}$ near $\delta=0$.
To accurately estimate the qubit frequency, we perform two-tone spectroscopy \cite{schuster_ac_2005, scarlino_all-microwave_2019}
(shown in Fig.~\ref{fig:suppfig_ac-stark}(c)) by applying a qubit pump tone at frequency $f_\mathrm{p}$ to the left plunger gate
(see Appendix~\ref{sec:app_setup}), while measuring the cavity at its resonance frequency $f_\mathrm{d}=f_\mathrm{c}$.
We fit a Lorentzian curve to the spectrum at each detuning value to extract $f_\mathrm{q}$ and fit it to the expression
$f_\mathrm{q}=\sqrt{\delta^2+4t_\mathrm{c}^2}/h$ yielding $t_\mathrm{c}/h = 1073 \pm 6.7$~MHz.
With this value of  $t_\mathrm{c}$, we fit the input-output model described in Appendix~\ref{sec:app_io} to the single tone spectroscopy data
shown in Fig.~\ref{fig:suppfig_ac-stark}(b) extracting a coupling strength of  $g_0/2\pi = 108.9 \pm 2.7$~MHz.

We then measure the ac Stark shift of the charge qubit at $\delta=0$ as a function of $P_\mathrm{VNA}$ for three cavity resonance frequencies
close to the ones used in Fig.~\ref{fig:suppfig_spectro}.
The dots in Fig.~\ref{fig:suppfig_ac-stark}(d) represent the measured qubit frequency shifts $\Delta\nu$
and the solid lines are linear fits according to Eq.~(\ref{eq:ac_stark}), from which we get the total attenuation factor $\beta$.
The extracted parameters for all three cavity frequencies are reported in Table~\ref{tab:app_ac_stark}.
For the measurements at $f_\mathrm{c}=2.978$~GHz, a different DQD configuration was used compared to $f_\mathrm{c}=3.644$~GHz and $f_\mathrm{c}=3.870$~GHz.
The same procedure is performed with the high-frequency setup in order to characterize the input loss
for the photon detection measurements at high frequencies (red dots in Fig.~\ref{fig:fig4}(a)).

\begin{table}[!ht]
    \begin{tabular}{rrrr}
        \toprule
 $f_\mathrm{c}$       & 2978~MHz        & 3644~MHz         & 3870~MHz            \\
        \midrule
 $\kappa/2\pi$             & 23.9 $\pm$ 0.2  & 26.4 $\pm$ 0.3   & 21.4 $\pm$ 0.2   \\
 $\kappa_\mathrm{c}/2\pi$  & 18.2 $\pm$ 0.1  & 22.2 $\pm$ 0.2   & 15.3 $\pm$ 0.1   \\
 $g_0/2\pi$                & 85.8 $\pm$ 1.8  & 108.9 $\pm$ 2.7  & 112.3 $\pm$ 3.0  \\
 $t_\mathrm{c}/h$ & 891.1 $\pm$ 9.4 & 1073.1 $\pm$ 6.7 & 1073.1 $\pm$ 6.7 \\
 $\beta\cdot10^9$     & 1.52 $\pm$ 0.13 & 0.62 $\pm$ 0.04  & 0.66 $\pm$ 0.05  \\
        \bottomrule
    \end{tabular}
    \caption{%
    Device parameters used for characterizing the input loss at the frequencies close to where the photon detection performance has been studied in the main text.
    $\kappa$ and $\kappa_\mathrm{c}$ are obtained from bare cavity spectroscopy measurements,
    $g_0$ ($t_\mathrm{c}$) is extracted from the single tone (two-tone) spectroscopy measurement reported in Fig.~\ref{fig:suppfig_ac-stark}(b) (Fig.~\ref{fig:suppfig_ac-stark}(c)).
    The total loss factor $\beta$ is extracted from the linear fit shown in Fig.~\ref{fig:suppfig_ac-stark}(d) using to Eq.~(\ref{eq:ac_stark}) and Eq.~(\ref{eq:nc}).
    All values except for the unitless $\beta$ are reported in MHz.
    }
    \label{tab:app_ac_stark}
\end{table}

\section{Cavity Characterization}\label{sec:app_kappas}

\begin{figure}[!ht]
    \includegraphics{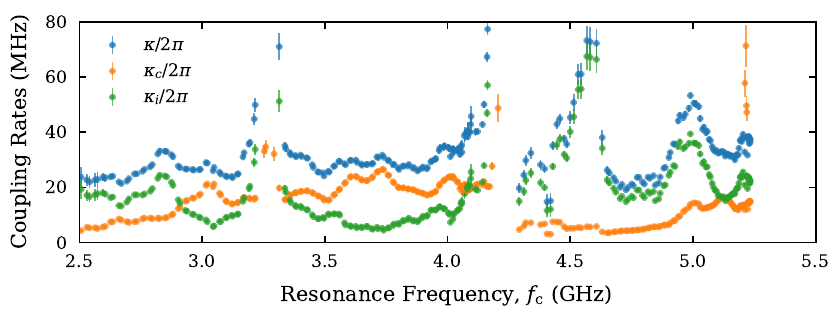}
	\caption{Cavity coupling rates as a function of cavity resonance frequency $f_\mathrm{c}$
            obtained from fitting the cavity magnetospectroscopy in Fig.~\ref{fig:fig1}(d) (for $f_\mathrm{c}<4.2$~GHz)
            and another similar measurement (for $f_\mathrm{c}>4.2$~GHz) using the low- and high-frequency setup, respectively
            (see Fig.~\ref{fig:suppfig_setup} in Appendix~\ref{sec:app_setup}).
	}
    \label{fig:suppfig_kappas}
\end{figure}

The bare cavity is characterized as a function of the cavity resonance frequency $f_\mathrm{c}$ with the DQD in Coulomb blockade, ensuring no interaction with the cavity.
The cavity-feedline coupling rate $\kappa_\mathrm{c}$, internal loss rate $\kappa_\mathrm{i}$, and total cavity loss rate $\kappa$ presented in Fig.~\ref{fig:suppfig_kappas}
are estimated by fitting the input-output model from Eq.(\ref{eq:io}) to magnetospectroscopy measurements
with a low cavity drive power $P_\mathrm{d}$ ($n_\mathrm{c}<1$).
For $f_\mathrm{c} < 4.2$~GHz, the data reported in Fig.~\ref{fig:fig1}(d), measured with the low-frequency circulator setup (described in Fig.~\ref{fig:suppfig_setup} in Appendix~\ref{sec:app_setup}), is used,
whereas the values for $f_\mathrm{c} > 4.2$~GHz are extracted from another measurement using the high-frequency circulator setup.

\begin{figure}[!ht]
    \includegraphics{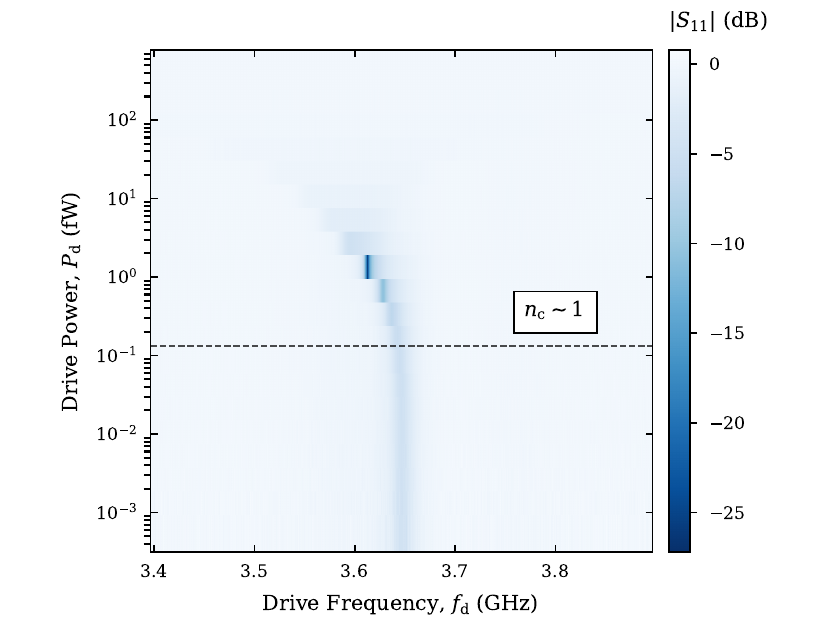}
	\caption{Normalized cavity reflectance $|S_{11}|$ as a function of the cavity drive frequency $f_\mathrm{d}$ and the feedline drive power $P_\mathrm{d}$
            with the cavity resonance frequency $f_\mathrm{c}\sim 3.646$~GHz.
            The power for which the average cavity photon number $n_\mathrm{c}\sim 1$ is indicated by a black dashed line.
            For higher drive power ($n_\mathrm{c}>1$) the resonance dip shifts to lower frequencies due to the self-Kerr non-linearity of the JJ array cavity \cite{krupko_kerr_2018}.
	}
    \label{fig:suppfig_pscan}
\end{figure}

The power dependence of the cavity response is studied by measuring the normalized cavity reflectance $|S_{11}|$
as a function of $f_\mathrm{d}$ and $P_\mathrm{d}$, and is reported in Fig.~\ref{fig:suppfig_pscan}.
The self-Kerr nonlinearity, arising from the intrinsic nonlinearity of the Josephson junctions,
leads to a power-dependent shift of $f_\mathrm{c}$ \cite{krupko_kerr_2018}.
Furthermore, the average cavity photon number $n_\mathrm{c}$ exhibits a nonlinear dependence on $P_\mathrm{d}$ at higher powers \cite{eichler_controlling_2014}.
These nonlinear effects do not impact the photodectector characterization carried out for $n_\mathrm{c}<1$
(below the dashed line in Fig.~\ref{fig:suppfig_pscan}).

\section{Impact of System Parameters on Efficiency}\label{sec:app_eff}

\begin{figure}[!ht]
    \includegraphics{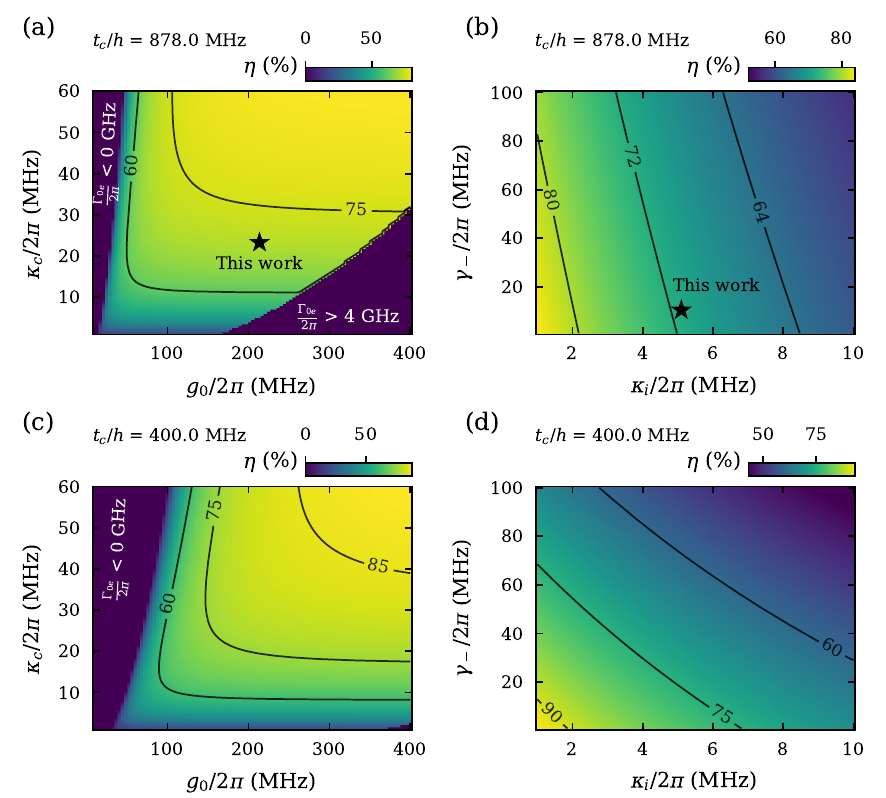}
	\caption{
    Dependence of the microwave photon detection efficiency $\eta$ on various system parameters.
    $\eta$ is calculated as a function of (a, c) $\kappa_\mathrm{c}$ and $g_0$, and (b, d) $\gammarelax$ and $\kappa_\mathrm{i}$.
    For panels (a) and (b), we use an inter-dot tunnel coupling strength of $t_\mathrm{c}/h \sim$ 878~MHz,
    as extracted from the configuration shown in Fig.~\ref{fig:fig2},
    whereas, for panels (c) and (d), we assume $t_\mathrm{c}/h = 400$~MHz.
    The relaxation rate of the qubit excited state to the reservoirs $\gammalead/2\pi$ is assumed to be tunable from 0--4~GHz
    to enforce $\kdqd = \kappa$ in all simulations. 
    The remaining system parameters, including $\kappa_\mathrm{i}$ and $g_0$, are assumed to be identical
    to the ones of the system configuration in Fig.~\ref{fig:fig2} and Fig.~\ref{fig:fig3}.
        }
    \label{fig:suppfig_efficiency}
\end{figure}

Figure~\ref{fig:suppfig_efficiency} demonstrates the microwave photon detection efficiency $\eta$ as a function of various parameters in the hybrid system.
In the following analysis, we assume that $\gammalead/2\pi$ is fully tunable within the range $0-4$~GHz to match
$\kdqd = 4g^2/\gammatot = \kappa$ with $\gammatot = \gammalead + \gammarelax + 2\gammadephasing$ at each point,
to ensure $\frac{4\kdqd\kappa}{(\kdqd + \kappa)^2} = 1$ (Eq.~(\ref{eq:photocurrent})).
In Figs.~\ref{fig:suppfig_efficiency}(a) and (c) $\eta$ is shown as a function of the cavity coupling strength $\kappa_\mathrm{c}$ and charge-photon coupling strength $g_0$. 
We also show $\eta$ as a function of the inter-dot relaxation rate $\gammarelax$ and internal cavity loss rate $\kappa_\mathrm{i}$
in Figs.~\ref{fig:suppfig_efficiency}(b) and (d).
For the analysis presented in Figs.~\ref{fig:suppfig_efficiency}(a) and (b), we use parameter values identical to those extracted from the experiments
at the cavity resonance frequency $f_\mathrm{c} = 3.645$~GHz
($\kappa/2\pi=28.1$~MHz, $\kappa_\mathrm{c}/2\pi=23.0$~MHz, $g_0/2\pi=213.7$~MHz, $t_\mathrm{c}/h=878.0$~MHz),
and assume symmetric QD-reservoir tunneling rates ($\Gamma_\mathrm{L}=\Gamma_\mathrm{R}$). 
For the inter-dot relaxation rate $\gammarelax$ and dephasing rate $\gammadephasing$, which cannot be individually assessed from the experiments,
we empirically assume $\gammarelax/2\pi = \gammadephasing/2\pi = 10$~MHz.
These parameters yield the calculated $\eta \sim 71 \%$ which is close to the measured $\eta \sim 67.7\%$ at $f_\mathrm{c} = 3.645$~GHz.

As clearly demonstrated from Fig.~\ref{fig:suppfig_efficiency}(a), a large $g_0$ allows high $\eta$ in a wide parameter window. 
Also, simply increasing $\kappa_\mathrm{c}$ by changing the device geometry, while keeping the other parameters constant, will allow $\eta$ to reach $\sim 80~\%$.
Alternatively as demonstrated in Fig.~\ref{fig:suppfig_efficiency}(b), decreasing the loss rate of each subsystem may also lead to $\eta > 80~\%$.
Figures~\ref{fig:suppfig_efficiency}(c) and (d) present calculations similar to the ones shown in Figs.~\ref{fig:suppfig_efficiency}(a) and (b),
but with smaller $t_\mathrm{c}/h = 400$~MHz while keeping the other parameters identical. 
This results in a larger directivity $\dir$ to yield higher $\eta$, where Fig.~\ref{fig:suppfig_efficiency}(d) explicitly demonstrates that further optimization of the system loss rates in addition to the smaller $t_\mathrm{c}$ can result in $\eta > 90 \%$.